\documentclass[12pt]{article}
\usepackage{a4wide}
\usepackage{amssymb}
\usepackage{graphicx}
\usepackage{caption}
\usepackage{subcaption}
\usepackage{mathdots}
\usepackage{color}
\usepackage[title, titletoc]{appendix}

\usepackage{a4wide}
\usepackage{amssymb}
\usepackage{graphicx}
\usepackage{caption}
\usepackage{subcaption}
\usepackage{mathdots}
\usepackage{slashed}
\usepackage{float}
\usepackage{bm}

\newcommand{\beq}{\begin{eqnarray}}
\newcommand{\eeq}{\end{eqnarray}}

\usepackage{blkarray}
\usepackage{multirow}

\usepackage{latexsym}
\usepackage{amsmath}

\usepackage{youngtab}

\def\drawbox#1#2{\hrule height#2pt 
        \hbox{\vrule width#2pt height#1pt \kern#1pt 
              \vrule width#2pt}
              \hrule height#2pt}

\def\Fund#1#2{\vcenter{\vbox{\drawbox{#1}{#2}}}}
\def\Asym#1#2{\vcenter{\vbox{\drawbox{#1}{#2}
              \kern-#2pt       
              \drawbox{#1}{#2}}}}

\def\funda{\Fund{6.5}{0.4}}
\def\asymm{\Asym{6.5}{0.4}}

\def\symm{\funda\kern-0.4pt\funda}

\def\makeatletter{\catcode`\@=11}
\makeatletter
\def\mathbox#1{\hbox{$\m@th#1$}}%
\def\math@ccstyles#1#2#3#4#5#6#7{{\leavevmode
      \setbox0\mathbox{#6#7}%
      \setbox2\mathbox{#4#5}%
      \dimen@ #3%
      \baselineskip\z@\lineskiplimit#1\lineskip\z@
      \vbox{\ialign{##\crcr
             \hfil \kern #2\box2 \hfil\crcr
             \noalign{\kern\dimen@}%
             \hfil\box0\hfil\crcr}}}}
\def\mathaccstyles{\math@ccstyles\maxdimen}
\def\maththroughstyles{\math@ccstyles{-\maxdimen}}
\def\unity%
 {\maththroughstyles{.45\ht0}\z@\displaystyle {\mathchar"006C}\displaystyle 1}

\def\be{\begin{eqnarray}}
\def\ee{\end{eqnarray}}

\begin{document}

\setcounter{table}{0}

\mbox{}
\vspace{2truecm}
\linespread{1.1}

\centerline{\LARGE \bf A note on D0-branes and instantons}

\vspace{10pt}

\centerline{\LARGE \bf  in 5d supersymmetric gauge theories}

\vspace{2truecm}

\centerline{
    {\large \bf Eran Avraham${}^{a}$} \footnote{eranav@post.bgu.ac.il}
{\bf and}
    {\large \bf Oren Bergman ${}^{a}$} \footnote{bergman@physics.technion.ac.il}}

\vspace{1cm}
\centerline{{\it ${}^a$ Department of Physics, Technion, Israel Institute of Technology}} \centerline{{\it Haifa, 32000, Israel}}
\vspace{1cm}

\centerline{\bf ABSTRACT}
\vspace{1truecm}

\noindent We refine a previous proposal for obtaining the multi-instanton partition function
from the supersymmetric index of the 1d supersymmetric gauge theory on the worldline of D0-branes.
We provide examples where the refinements are crucial for obtaining the correct result.

\newpage

\tableofcontents

\section{Introduction}

The problem of counting instantons in supersymmetric gauge theories with eight supersymmetries has received some recent interest due to its appearance in the study of five-dimensional and six-dimensional superconformal field theories.
In many cases these theories correspond to UV fixed points of 5d ${\cal N}=1$ supersymmetric gauge theories,
in which the instantons, which are particles in five dimensions, provide a crucial ingredient in identifying the properties of 
the UV theory. 
To facilitate this we must be able to correctly compute the charges and degeneracies of the instanton states, namely to ``count" instantons.
In order to do this we must quantize the multi-instanton moduli space as provided by the ADHM construction \cite{Atiyah:1978ri}.
In general this is a difficult problem since the moduli space contains singularities corresponding to instantons of vanishing size.
Another way to say this is that the corresponding supersymmetric linear sigma model \cite{Witten:1994tz} is not well defined in the UV,
and requires a UV completion.

There have been a number of approaches for dealing with these singularities.
Originally, Nekrasov used the so-called Omega-deformation to remove the singularities and compute the
instanton partition function \cite{Nekrasov:2002qd}. However this approach only works for $U(N)$.
Subsequently, Nekrasov and Shadchin were able to overcome the singularities for other gauge groups 
by lifting to five dimensions \cite{Nekrasov:2004vw,Shadchin:2005mx}.

An alternative approach is to find a UV completion of the ADHM sigma model, and attempt to extract from it the relevant data
of the instanton moduli space.
This approach is strongly motivated by string theory, in which five-dimensional supersymmetric gauge theories
may be realized in terms of D4-branes in Type IIA superstring theory, and the instantons correspond to D0-branes
inside the D4-branes  \cite{Douglas:1996uz}.
The ADHM sigma model is naturally embedded in the 1d ${\cal N}=(0,4)$ supersymmetric gauge theory on the D0-brane worldline.
It is the low energy effective theory on the Higgs branch of the gauge theory.
On the other hand the 1d ${\cal N}=(0,4)$ gauge theory is a UV complete theory.

Recent advances in supersymmetric localization have made it possible to compute the exact partition function, or index,
of such theories \cite{Kim:2011mv,Kim:2012gu,Hwang:2014uwa,Hwang:2016gfw}.
The multi-instanton partition function is then naively given by summing over instanton sectors:
\be
\label{QMindex1}
Z_{QM} = \sum_{k=0}^\infty I_k \, q^k \,,
\ee
where $I_k$ is the index of the 1d ${\cal N}=(0,4)$ gauge theory of $k$ instantons, and $q$ is the instanton fugacity.
The index $I_k$ is computed using supersymmetric localization, which reduces the path integral to ordinary contour 
integrals over the complexified holonomies of the gauge multiplet.
The general form of the multi-particle index is given by the plethystic exponent of the single-particle index $f(q)$ (which will in general get contributions from multi-instanton bound states):
\be
\label{QMindex2}
Z_{QM} = \mbox{PE}[f(q)] = \exp\left[\sum_{n=1}^\infty \frac{f(q^n)}{n} \right].
\ee
By comparing with (\ref{QMindex1}), one can extract information about the spectrum of one-particle states, 
up to a given instanton number $k$.
Indeed one can often guess the complete $f(q)$ by computing $I_k$ for a small number of $k$'s.
This was done for the ${\cal N}=2$ supersymmetric 5d $U(N)$ gauge theory in \cite{Kim:2011mv},
and used to show the existence of instanton bound states at all instanton numbers, as required
by the conjecture that this theory flows to the 6d A-type $(2,0)$ theory in the UV.

However, more generally one has to overcome the following difficulty.
The moduli space of the 1d gauge theory generically also has additional branches that contribute to the index.
For ${\cal N}=2$ theories the 1d gauge theory has a Coulomb branch,
and for ${\cal N}=1$ theories it can have both a Coulomb branch and a twisted Higgs branch parameterized by
a twisted hypermultiplet.
In the brane description these branches correspond to moving the D0-branes out of and away from the D4-branes.
The contribution of the states on the extra branches must be removed in order to obtain the instanton partition function,
which is just the contribution of the states on the Higgs branch.
It seems natural to conjecture that the contribution of the extra states is accounted for by the {\em reduced} 1d gauge theory of the D0-branes
in the absence of the D4-branes, namely
\be
\label{Zextra1}
Z_{extra} = Z_{QM,red} \,,
\ee 
where $Z_{QM,red}$ is computed for the 1d gauge theory without the D4-branes, and
\be
\label{InstantonIndex1}
Z_{inst} = \frac{Z_{QM}}{Z_{extra}} \,.
\ee
This was originally conjectured in \cite{Kim:2012gu} for the class of ${\cal N}=1$ theories with gauge group
$Sp(N)$, $N_F\leq 7$ flavor hypermultiplets, and an antisymmetric hypermultiplet, that are realized
in terms of D4-branes in the background of an orientifold 8-plane and $N_f$ D8-branes \cite{Seiberg:1996bd}.
It was shown in \cite{Kim:2012gu}, by computing the 1-instanton contribution, 
that this correctly reveals the enhanced exceptional global symmetries for $N_F\leq 5$.
Subsequently $Z_{extra}$ was computed to higher orders in these theories and an all-orders expression was conjectured in 
\cite{Hwang:2014uwa}. In particular this completed the work of \cite{Kim:2012gu} and showed
the enhanced $E_7$ and $E_8$ symmetries for $N_F=6$ and $N_F=7$, respectively.

The same issue was addressed  in the ${\cal N}=2$ theories realized in terms of D4-branes in flat space or in the background
of an orientifold 4-plane in \cite{Hwang:2016gfw}.
For $U(N)$ it was shown that $Z_{extra}=1$, in line with the result of \cite{Kim:2011mv}, which did not include this correction.
However for $O(N)$ and $Sp(N)$ the contribution of the extra branches is non-trivial.
Furthermore, for $Sp(N)$ it was found that an additional correction is required;
in this case $Z_{extra}$ overcounts by a multiplicitive quantity, 
which corresponds to bound states of D0-branes that are confined to the orientifold 4-plane.
This last observation can be understood from the fact that the reduced 1d gauge theory in this case still has a remnant of the 
original Higgs branch, whose states we do not want to remove.
The additional correction adds these contributions back in.

We would like to formalize this into the following modified proposal 
for the contribution of the extra states:
\be
\label{InstantonIndex2}
Z_{extra} = \frac{Z_{QM,red}}{\mbox{PE}[f_{D0_c}(q)]}
\ee
where 
$f_{D0_c}(q)$ is the single-particle index of D0-branes that are confined to any 5d defect that remains after the 4-branes are removed.
In addition, instead of simply removing the D4-branes by setting $N=0$ in the 1d gauge theory, the reduced theory
should be understood as the low-energy effective theory that results by turning on a mass for the fields
charged under the 4-brane gauge symmetry and integrating them out.
Geometrically this corresponds to moving the 4-branes away from the 0-branes.
Depending on the theory, this can shift the values of Chern-Simons (CS) or discrete theta parameters.
In all the examples studied in \cite{Kim:2011mv,Kim:2012gu,Hwang:2014uwa,Hwang:2016gfw} the net shifts vanish,
but, as we will see, in general they do not.

In this note we will present additional examples of 5d ${\cal N}=1$ gauge theories in support of our conjecture (\ref{InstantonIndex2}).
We will consider the theories obtained by a $\mathbb{Z}_2$  orbifold of the brane configuration of the ${\cal N}=1$ $Sp(N)+\asymm$ theory \cite{Bergman:2012kr}.
There is a choice in this orbifold corresponding to the action of worldsheet parity in the twisted sector.
The so-called orbifold with vector structure leads to the quiver theory with $Sp(N_1)\times Sp(N_2)$ and a bi-fundamental
hypermultiplet, as well as flavors charged under either gauge group.
The orbifold without vector structure gives an $SU(N)$ gauge group with two antisymmetric hypermultiplets, plus the flavors.
As we will see, these two examples exhibit the need both for the second correction factor in (\ref{InstantonIndex2}),
as well as for the shift in the CS or discrete theta parameters in the reduced theory. 

The outline for the rest of the paper is as follows.
In section 2 we will review the general structure of 1d ${\cal N}=(0,4)$ gauge theories.
In section 3 we will review the results for $Z_{extra}$ in the 
5d ${\cal N}=2$ theories and ${\cal N}=1$ $Sp(N)+\asymm$ theory, and show that they are consistent with our proposal.
In sections 4 and 5 we will apply our proposal to the $\mathbb{Z}_2$ orbifold theories, and perform some basic checks.
Section 6 contains our conclusions.

\section{General structure of ${\cal N}=(0,4)$ supersymmetric QM}

A 1d theory with ${\cal N}=(0,4)$ supersymmetry has four supercharges $Q^{\dot{\alpha}A}$ transforming in 
the $({\bf 2},{\bf 2})$ representation of $SO(4)_R = SU(2)_r\times SU(2)_{r'}$.\footnote{We follow the conventions of 
\cite{Hwang:2014uwa}, and use a dotted greek index for 
$SU(2)_r$ and an undotted capital latin index for $SU(2)_{r'}$. The latter is denoted $SU(2)_R$ in \cite{Hwang:2014uwa}.
We reserve subscripts $R$ and $L$ for fermion spin.}
This kind of supersymmetry admits four types of supermultiplets:
a vector multiplet $(A_0, \varphi, \lambda^{\dot{\alpha}A})$, a hypermultiplet $(\phi_{\dot{\alpha}},\psi_A)$, 
a twisted hypermultiplet $(\phi_A,\psi_{\dot{\alpha}})$, and a Fermi multiplet $\psi$.
In terms of ${\cal N} = (0,2)$ multiplets, the $(0,4)$ vector multiplet decomposes into a vector multiplet and two Fermi multiplets,
and the hypermultiplet and twisted hypermultiplet each decompose into a chiral and an anti-chiral multiplet.

The ${\cal N} = (0,4)$ theory for instantons in a 5d ${\cal N}=1$ gauge theory also has an $SU(2)_\ell$ global symmetry, 
such that $SU(2)_\ell \times SU(2)_r$ is the rotation symmetry of the 5d theory. 
Furthermore, for 5d gauge theories that are realized on D4-branes, possibly in the presence of D8 ``flavor branes", 
the 1d (and 5d) theory has an additional $SU(2)_{\ell'}$ global symmetry (denoted $SU(2)_L$ in \cite{Hwang:2014uwa}).
The symmetry $SU(2)_{\ell'}\times SU(2)_{r'}$ corresponds to rotations in the directions transverse to the D4-branes and along the D8-branes.
Table~\ref{ADHMQM} summarizes the generic content of the theory in these cases.
The subscripts $R$ and $L$ on the fermions denote their chirality in the lift to two dimensions.
In the one-dimensional theory this corresponds to the spin.
The top three rows show the 1d fields originating from the 5d gauge multiplet.
The rest are associated with the 5d matter hypermultiplets.
A 5d matter field in the fundamental representation gives a 1d Fermi multiplet $\xi$.
Matter fields in higher tensor representations lead to the 1d fields shown in the bottom three rows.
The four scalar fields in $X$ correspond to the positions of the D0-branes along the D4-branes, and the four scalar
fields in $Y$ together with the vector multiplet scalar $\varphi$ correspond to the positions of the D0-branes transverse to the D4-branes.

The schematic form of the scalar potential is given by (see for example \cite{Aharony:1997pm,Tong:2014yna,Tong:2014cha})
\beq
\label{ScalarPotential1}
V= \text{Tr}\left[([X, X]+q^2)^2+[Y, Y]^2+[X,Y]^2 +[X,\varphi]^2 + [Y,\varphi]^2 + q^2(Y^2 + \varphi^2)\right] \,.
\eeq
The theory generically has three distinct branches of vacua:
\begin{itemize}
\item A Higgs branch given by $\varphi = Y = 0$ and $[X,X]+q^2 = 0$
\item A Coulomb branch given by $q=0$ and $[X,X]=[Y,Y]=[X,Y]=[X,\varphi]=[Y,\varphi]=0$
\item A twisted Higgs branch given by $\varphi = q = 0$ and $[X,X]=[Y,Y]=[X,Y]=0$
\end{itemize}
There are also mixed branches, where some components of $\varphi$ vanish and some components of $q$ vanish.
The reduced theory obtained by removing the fields $q$, $\psi_L$ and $\psi_R$ will clearly have the same Coulomb and twisted Higgs branches
as the original theory.
The main issue will be whether it also possesses a remnant of the original Higgs branch, namely whether the space defined by
$\varphi = Y = [X,X]=0$ has components not contained in the Coulomb or twisted Higgs branches.

\begin{table}[h!]
\begin{center}
\begin{tabular}{|l|l|c|c|c|c|c|}
  \hline 
  $(0,4)$ multiplet & Fields & $SU(2)_\ell\times SU(2)_r$  & $SU(2)_{\ell'}\times SU(2)_{r'}$ & $G_{D0}$ & $G_{D4}$ & $G_{D8}$ \\
 \hline 
 Vector&$A_0, \varphi$ & $({\bf 1},{\bf 1})$ & $({\bf 1},{\bf 1})$ &  $\mbox{adj}$ &  ${\bf 1}$ & ${\bf 1}$\\
 &$\lambda_R$ & $({\bf 1},{\bf 2})$ & $({\bf 1},{\bf 2})$ &  & & \\
 \hline
Hyper (real) &$X$ & $({\bf 2},{\bf 2})$ & $({\bf 1},{\bf 1})$ & $R_X$ & ${\bf 1}$ & ${\bf 1}$\\
& $\chi_L$ & $({\bf 2},{\bf 1})$ & $({\bf 1},{\bf 2})$ & & &  \\
 \hline
 Hyper (complex) &$q$ & $({\bf 1},{\bf 2})$ &  $({\bf 1},{\bf 1})$ & $\funda$ & $\bar{\funda}$ & ${\bf 1}$ \\
 &$\psi_L$ &  $({\bf 1},{\bf 1})$ & $({\bf 1},{\bf 2})$  & &  & \\
 \hline\hline
 Fermi & $\xi_R$ & $({\bf 1},{\bf 1})$ & $({\bf 1},{\bf 1})$ & $\funda$ & ${\bf 1}$ & $\bar{\funda}$\\
 \hline\hline
Twisted Hyper &$Y$ & $({\bf 1},{\bf 1})$ & $({\bf 2},{\bf 2})$ & $R_Y$  & ${\bf 1}$ & ${\bf 1}$\\ 
& $\lambda_L$ & $({\bf 1},{\bf 2})$ & $({\bf 2},{\bf 1})$ &  & & \\
 \hline
 Fermi & $\chi_R$ & $({\bf 2},{\bf 1})$ & $({\bf 2},{\bf 1})$ & $R_{\chi}$  & ${\bf 1}$ & ${\bf 1}$\\
  \hline
Fermi & $\psi_R$ &  $({\bf 1},{\bf 1})$ &  $({\bf 2},{\bf 1})$ & $\funda$ & $R_{\psi}$ & ${\bf 1}$\\
 \hline
 \end{tabular}
 \end{center}
\caption{Spectrum of 1d ${\cal N}=(0,4)$ gauge theory in a generic D0-D4-D8 system.
The unspecified representations depend on the specific model being considered.
The subscript $L,R$ refers to the spin of the corresponding fermion.}
\label{ADHMQM}
\end{table}

\section{Comparison with known results}

\subsection{The ${\cal N}=2$ theories}

The 5d ${\cal N}=2$ theories are completely characterized by the gauge group $G_{D4} = U(N)$, $O(N)$, 
or $Sp(N)$.
In the ${\cal N}=1$ language these theories have a single matter hypermultiplet in the adjoint representation. 
The $U(N)$ theory is realized on $N$ parallel D4-branes in Type IIA string theory.
Adding an $\mbox{O4}^-$ or $\widetilde{\mbox{O4}}^-$ plane gives the $O(N)$ theory with $N$ even or odd, respectively.
Adding an $\mbox{O4}^+$ or $\widetilde{\mbox{O4}}^+$ plane gives the $Sp(N)$ theory with the discrete theta parameter 
$\theta = 0$ or $\pi$, respectively \cite{Tachikawa:2011ch}.
These theories have six-dimensional UV fixed points corresponding to 6d A-type or D-type $(2,0)$ theories
described by M5-branes in M theory.

\begin{figure}[h]
\center
\includegraphics[height=0.15\textwidth]{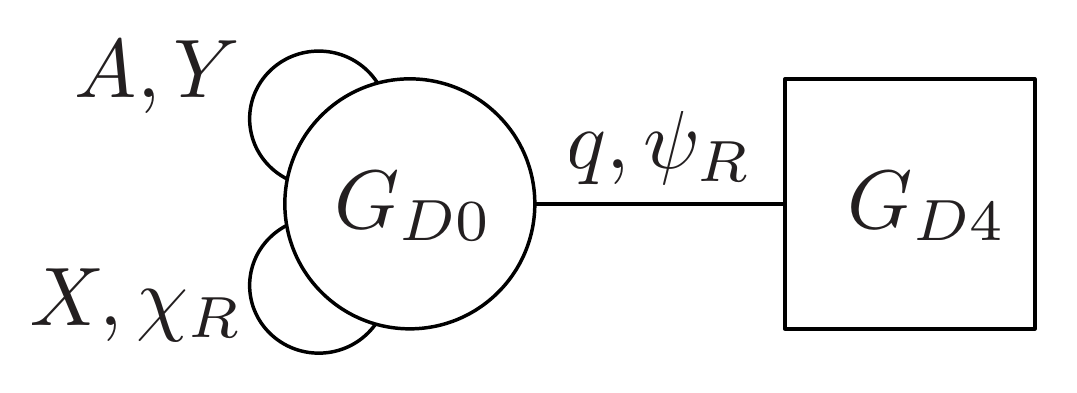} 
\caption{1d ${\cal N}=(4,4)$ theories for the 5d ${\cal N}=2$ theories.}
\label{UNwithAdjoint}
\end{figure}

The D0-brane gauge theory in these cases is shown in Fig.~\ref{UNwithAdjoint},
with $G_{D0} = U(k), Sp(k)$, and $O(k)$ when
$G_{D4} = U(N)$, $O(N)$, and $Sp(N)$, respectively.\footnote{For the case of $G_{D0}=U(k)$ and $G_{D4}=U(N)$,
the global symmetry of the 1d $U(k)$ gauge theory is actually $SU(N)$, since the $U(1)\in U(N)$ is gauged.}
The twisted hypermultiplet $Y$ transforms in the adjoint representation of $G_{D0}$, 
the multiplets $q$ and $\psi_R$ transform in the bifundamental of $G_{D0}\times G_{D4}$, 
and $X$ and $\chi_R$ transform in the adjoint, antisymmetric, and symmetric representation of $G_{D0}$ in the three cases, respectively.
These theories really have ${\cal N} = (4,4)$ supersymmetry: the vector multiplet and twisted hypermultiplet combine into a $(4,4)$ vector multiplet, 
and each hypermultiplet combines with a Fermi multiplet into a $(4,4)$ hypermultiplet.
There is a Coulomb branch that combines the ${\cal N}=(0,4)$ Coulomb and twisted Higgs branches parameterized by $(\varphi, Y)$, 
and a Higgs branch parameterized by $(X,q)$.

The reduced 1d gauge theories are given by removing the fields $\{q,\psi_L\}$ and $\psi_R$.
The multi-particle indices corresponding to the reduced theories in the different cases were given in \cite{Hwang:2016gfw} as follows
(the notation is reviewed in the Appendix):
\be
\label{N=2ReducedIndex}
Z_{QM,red} = \left\{
\begin{array}{ll}
1 & U(N) \\[10pt]
\mbox{PE}\left[\frac{t^2 (t + \frac{1}{t})  (v + \frac{1}{v} - u - \frac{1}{u})}{2(1-tu)(1-\frac{t}{u})(1+tv)(1+\frac{t}{v})}\, 
\frac{q}{1-q}\right]  & O(N), \; N \; \mbox{even} \\[10pt]
\mbox{PE}\left[\frac{t^2 (t + \frac{1}{t}) (v + \frac{1}{v} - u - \frac{1}{u})}{2(1-tu)(1-\frac{t}{u})(1+tv)(1+\frac{t}{v})}\, 
\frac{-q}{1+q}\right] & O(N), \; N \; \mbox{odd}\\[10pt]
\mbox{PE}\left[\frac{t^2  (t + \frac{1}{t}) (v + \frac{1}{v} - u - \frac{1}{u})}
{2(1-tu)(1-\frac{t}{u})(1+tv)(1+\frac{t}{v})}\, \frac{q^2}{1-q^2}
+ \frac{t(v + \frac{1}{v} - u - \frac{1}{u})}{(1-tu)(1-\frac{t}{u})}\, \frac{q}{1-q^2} \right] & Sp(N)_0 \\[10pt]
\mbox{PE}\left[\frac{t^2 (t + \frac{1}{t}) (v + \frac{1}{v} - u - \frac{1}{u})}
{2(1-tu)(1-\frac{t}{u})(1+tv)(1+\frac{t}{v})}\, \frac{-q^2}{1+q^2}
+ \frac{t(v + \frac{1}{v} - u - \frac{1}{u})}{(1-tu)(1-\frac{t}{u})}\, \frac{q^2}{1-q^4} \right] & Sp(N)_\pi
\end{array}
\right.
\ee
However it was argued in \cite{Hwang:2016gfw} that the second term in the one-particle index 
in the plethystic exponent for the $Sp(N)$ theories should be discarded to agree with the M theory viewpoint.
The contribution of the extra states was therefore claimed to be given by
\be
\label{N=2ExtraIndex}
Z_{extra} = \left\{
\begin{array}{ll}
Z_{QM,red} & U(N), O(N) \\[10pt]
Z_{QM,red} \cdot \left(\mbox{PE}\left[\frac{t(v + \frac{1}{v} - u - \frac{1}{u})}{(1-tu)(1-\frac{t}{u})}\, \frac{q}{1-q^2} \right]\right)^{-1} & Sp(N)_0 \\[10pt]
Z_{QM,red}  \cdot \left(\mbox{PE}\left[\frac{t(v + \frac{1}{v} - u - \frac{1}{u})}{(1-tu)(1-\frac{t}{u})}\, \frac{q^2}{1-q^4} \right]\right)^{-1} & Sp(N)_\pi
\end{array}
\right.
\ee
As we will now demonstrate these identifications are consistent with our proposal in (\ref{InstantonIndex2}).

Our proposal has two new ingredients.
First, one must take into account the possible shifts in the values of the gauge theory parameters as a result of integrating
out rather than just removing the massive fermions $\psi_L$ and $\psi_R$.
This is relevant for the 1d $U(k)$ and $O(k)$ theories.

The 1d $U(k)$ theory admits a CS term $\kappa\int \mbox{Tr}A$, with $\kappa\in\mathbb{Z}$, that is inhereted from the CS term of
the 5d $U(N)$ theory (see for example \cite{Collie:2008vc}). 
This can be thought of as a background $U(1)$ gauge charge of size $\kappa$.
The effective CS parameter of the reduced theory will get a contribution from the massive fermions that are integrated out.
Each fermion in the fundamental representation of $U(k)$ contributes $\frac{1}{2}\mbox{sign}(m)$,
where sign of the mass is given by the product of the spin and the diagonal $U(1)$ charge of the fermion.
The net shift vanishes since the hypermultiplet fermion $\psi_L$ and the Fermi multiplet fermion $\psi_R$ have the same charge but 
opposite spin.

The 1d $O(k)$ theory admits a discrete theta parameter $\theta$ taking values in $\{0,\pi\}$, that is inhereted from
the discrete theta parameter of the 5d $Sp(N)$ theory \cite{Bergman:2013ala}.
This can be regarded as a background $O(1) = \mathbb{Z}_2$ gauge charge.
The effective theta parameter of the reduced theory will get
a contribution depending on the sign of the determinant of the fermion mass matrix
\cite{Intriligator:1997pq}. The mass-deformation moving the 4-branes breaks $Sp(N)$ to $U(N)$ .
The fermions $\psi_R$ and $\psi_L$ each decompose into $N$ states carrying positive charge under the diagonal $U(1)$, and $N$
carrying negative charge. Therefore there are $2N$ fermions of negative mass and $2N$ of positive mass, and therefore no shift in $\theta$.

The second ingredient of our proposal 
has to do with the overcounting of extra states when the moduli space of 
the reduced theory contains a remnant of the Higgs branch of the original theory.
Equivalently, the question is whether the reduced theory has a Higgs branch that is separate from its Coulomb branch.
The would-be Higgs branch of the reduced theory is given by $\varphi_H = Y_H = 0$ and
\be
\label{ReducedHiggsBranch}
X_H = 
\left\{
\begin{array}{ll}
\mbox{diag}[x_1,\ldots,x_k] & U(k) \\
\mbox{diag}[x_1 \sigma_2,\ldots,x_k \sigma_2] & Sp(k) \\
\mbox{diag}[x_1,\ldots,x_{2n}]  & O(k=2n) \\
\mbox{diag}[x_1,\ldots,x_{2n+1}]  & O(k=2n+1) \,.
\end{array}
\right.
\ee
On the other hand on the Coulomb branch
\be
\label{CoulombBranchPhi}
\varphi_C = 
\left\{
\begin{array}{ll}
\mbox{diag}[a_1,\ldots,a_k] & U(k) \\
\mbox{diag}[a_1 \mathbb{I},\ldots,a_k\mathbb{I}]  & Sp(k) \\
\mbox{diag}[a_1 \sigma_2,\ldots,a_n \sigma_2] & O(2n) \\
\mbox{diag}[a_1 \sigma_2,\ldots,a_n \sigma_2 ,0] & O(2n+1) 
\end{array}
\right.
\ee
with $Y_C$ having the same form, and
\be
\label{CoulombBranchX}
X_C = 
\left\{
\begin{array}{ll}
\mbox{diag}[x_1,\ldots,x_k] & U(k) \\
\mbox{diag}[x_1 \sigma_2,\ldots,x_k \sigma_2] & Sp(k) \\
\mbox{diag}[x_1 \mathbb{I},\ldots,x_n\mathbb{I}]  & O(2n) \\
\mbox{diag}[x_1 \mathbb{I},\ldots,x_n\mathbb{I},x] & O(2n+1) \,. 
\end{array}
\right.
\ee
We see that only the orthogonal theories have a separate Higgs branch.
This can be understood as the possibility for a bulk D0-brane to split into a pair of fractional
D0-branes confined to an $\mbox{O4}^+$-plane.
The split phase, in which the two fractional D0-branes are restricted to moving on the orientifold plane, 
is a remnant of the original Higgs branch, in which the D0-branes were inside the D4-branes.
This splitting is not possible on an $\mbox{O4}^-$-plane, which is why there is no Higgs branch
in the reduced $Sp(k)$ theory.

Our proposal is therefore consistent with the identification of $Z_{extra}$  for the 5d $U(N)$ and $O(N)$ theories
in (\ref{N=2ExtraIndex}).
For the $Sp(N)$ theory we need to divide by the correction factor corresponding to confined D0-branes, $\mbox{PE}[f_{D0_c}(q)]$.
The quantity $f_{D0_c}(q)$ is the one-particle index for D0-branes confined to the $\mbox{O4}^+$-plane.
This quantity was originally studied in  \cite{Keurentjes:2002dc}.
The ${\cal O}(q)$ term is just the index of the reduced 1d theory with $k=1$, so
\be
\label{FractionalD0brane}
f_{D0_c}(q) =  \frac{1}{2}(1+ e^{i\theta}) I_1^X I_1^{\chi_R} \, q + {\cal O}(q^2)= \frac{1}{2}(1+ e^{i\theta}) \,
\frac{t(u + \frac{1}{u} - v - \frac{1}{v})}{(1-tu)(1-\frac{t}{u})} \, q + {\cal O}(q^2) \,.
\ee
This reproduces ${\cal O}(q)$ term in the correction factor for $\theta=0$ in (\ref{N=2ExtraIndex}).
The higher order terms are more difficult to obtain.
For $\theta = \pi$ the single D0-brane state is projected out due to the background $\mathbb{Z}_2$ charge,
and the lowest order contribution is at ${\cal O}(q^2)$, in agreement with (\ref{N=2ExtraIndex}).
This corresponds to a state with two confined D0-branes connected by a string which cancels the background charge.

\subsection{The $Sp(N) + \asymm$ theory} 

This is the theory on $N$ D4-branes in an $\mbox{O8}^-$-plane with $N_f$ D8-branes.
For $N_f \leq 7$ this theory has a 5d UV fixed point with an enhanced $E_{N_f +1}$ global symmetry \cite{Seiberg:1996bd}.
For $N_f = 8$ it corresponds to the 6d $(1,0)$ E-string theory.
For $N_f=0$ there is another possibility associated to the choice of discrete theta parameter, the so-called
$\tilde{E}_1$ theory, in which the global $U(1)$ symmetry is not enhanced \cite{Morrison:1996xf,Douglas:1996xp,Intriligator:1997pq}.
For $N_f>0$ the theta parameter is not physical, since it can be removed by a transformation in the parity-reversing component of 
the global $O(2N_f)$ symmetry.

\begin{figure}[h]
\center
\includegraphics[height=0.2\textwidth]{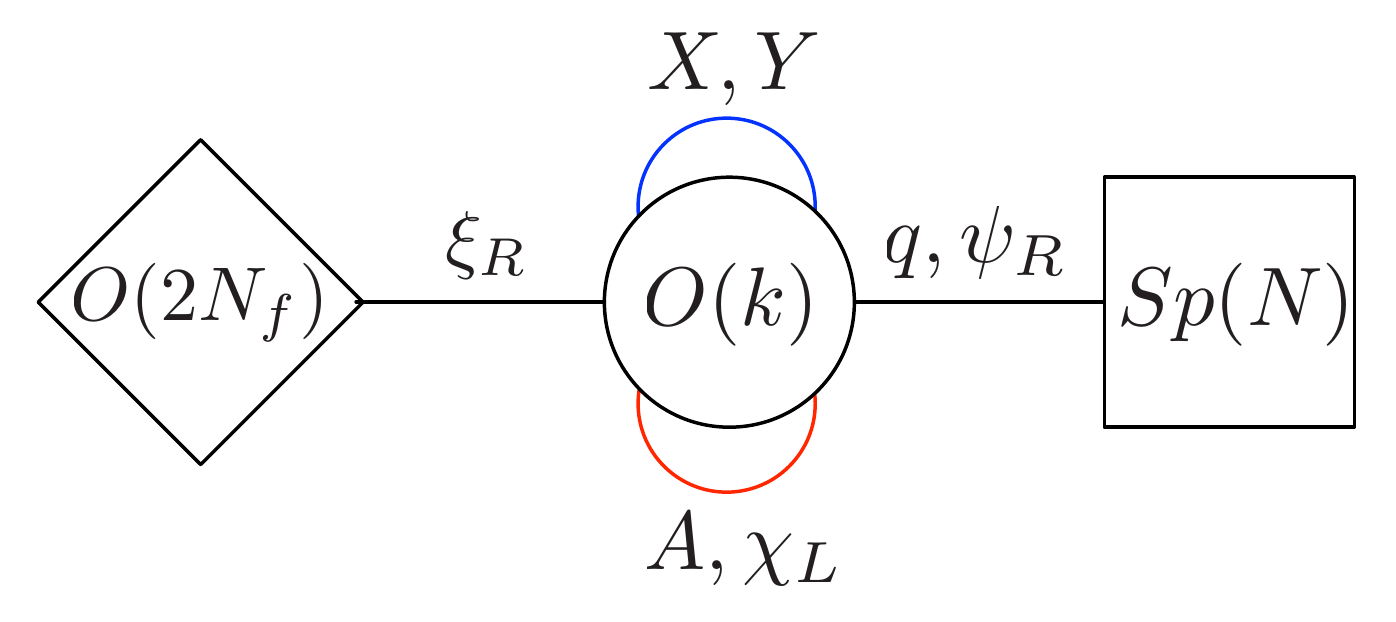} 
\caption{The 1d ${\cal N}=(0,4)$ theory for the 5d $Sp(N)$ with an antisymmetric hypermultiplet and $N_f$ fundamental
hypermultiplets.}
\label{SpNwithAntisymmetric}
\end{figure}

The corresponding 1d ${\cal N}=(0,4)$ gauge theory  on the D0-branes is shown in Fig.~\ref{SpNwithAntisymmetric}.
The hypermultiplet $X$ and twisted hypermutiplet $Y$ both transform in the symmetric representation of $O(k)$.
The reduced 1d gauge theory is again given by removing the fields $\{q,\psi_L\}$ and $\psi_R$.
The multi-particle indices corresponding to the reduced theory with $0\leq N_f \leq 8$ were given in \cite{Hwang:2014uwa} as follows:
\be
\label{SpNReducedIndex}
Z_{QM,red} = \left\{
\begin{array}{ll}
 \mbox{PE}\left[\frac{-t^2}{(1-tu)(1-\frac{t}{u})(1-tv)(1-\frac{t}{v})} \, q \right]  & N_f = 0 \\[10pt]
  \mbox{PE}\left[\frac{-t^2}{(1-tu)(1-\frac{t}{u})(1-tv)(1-\frac{t}{v})} \, 
q \chi[f]^{O(2N_f)}_{{\bf 2}^{N_f -1}} \right] & N_f = 1, \ldots , 5 \\[10pt]
\mbox{PE}\left[\frac{-t^2}{(1-tu)(1-\frac{t}{u})(1-tv)(1-\frac{t}{v})} 
\left(q \chi[f]^{O(12)}_{{\bf 32}} + q^2 \right) \right] & N_f = 6 \\[10pt]
\mbox{PE}\left[\frac{-t^2}{(1-tu)(1-\frac{t}{u})(1-tv)(1-\frac{t}{v})} 
\left(q \chi[f]^{O(14)}_{{\bf 64}} + q^2 \chi[f]^{O(14)}_{{\bf 14}}\right) \right] & N_f = 7\\[10pt]
\mbox{PE}\Bigg[\frac{-(t+t^3)(u + \frac{1}{u} + v + \frac{1}{v})}{2(1-tu)(1-\frac{t}{u})(1-tv)(1-\frac{t}{v})} 
\, \frac{q^2}{1-q^2}  & \\
 \mbox{} - \frac{t^2}{(1-tu)(1-\frac{t}{u})(1-tv)(1-\frac{t}{v})} \left( 
\chi[f]^{O(16)}_{{\bf 120}} \frac{q^2}{1-q^2} + \chi[f]^{O(16)}_{{\bf 128}} \frac{q}{1-q^2}\right)   \Bigg] & N_f = 8
\end{array}
\right.
\ee
where $f$ denotes collectively the flavor fugacities.
In all cases these were claimed to give precisely the extra factors, namely
\be
 Z_{extra} = Z_{QM,red} \,.
 \ee
As a slight generalization we also give the result for $N_f=0$ and $\theta = \pi$ (which we checked to ${\cal O}(q^4)$):
\be
Z_{extra} [N_f = 0, \theta = \pi]=Z_{QM,red}[N_f = 0, \theta = \pi] = 1 \,.
\ee

These results are also consistent with our proposal.
First we observe that the net shift in the discrete theta parameter in the $N_f=0$ theory vanishes 
since there are an even number $2N$ of negative mass fermions.
As to the moduli space of the reduced theory, on the Coulomb branch $\varphi_C$ has the same form as in  (\ref{CoulombBranchPhi}) for the orthogonal theory,
and $X_C$ and $Y_C$ take the same form as in (\ref{CoulombBranchX}) for the orthogonal theory. 
There is also a twisted Higgs branch, on which $\varphi_{TH}= 0$ and 
\be
Y_{TH} &=& \mbox{diag}[y_1,\ldots,y_k] \nonumber \\
X_{TH} &=& \mbox{diag}[x_1,\ldots,x_k] \,.
\ee
The would-be Higgs branch of the reduced theory is given by $\varphi_H = Y_H = 0$ and $X_H =  \mbox{diag}[x_1,\ldots,x_k]$,
and is therefore completely contained in the twisted Higgs branch.
There is no separate Higgs branch in this case, and therefore no correction factor.
The reduced theory precisely accounts for the extra states.

\section{The $SU(N) + 2 \, \asymm$ theory}

\begin{figure}[h]
\center
\includegraphics[height=0.2\textwidth]{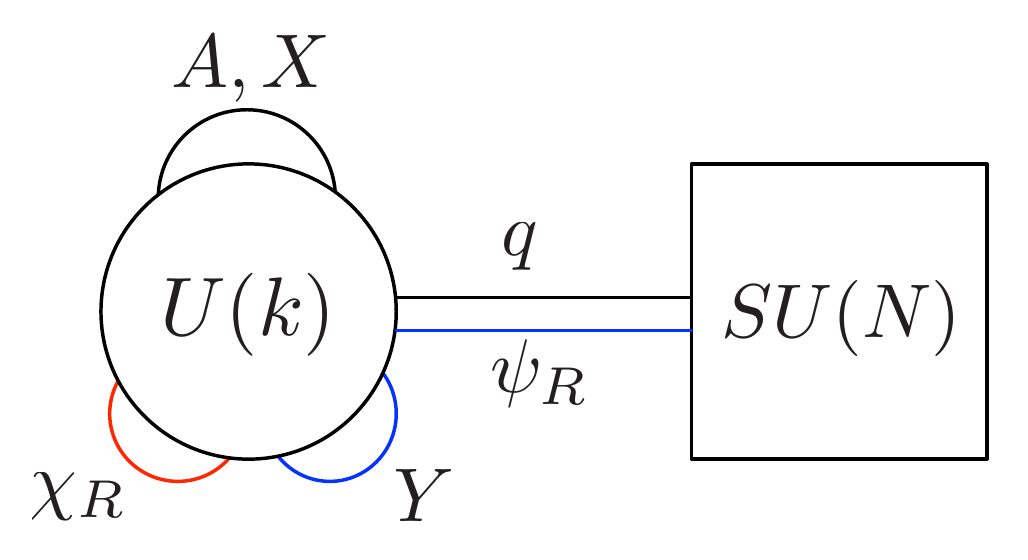} 
\caption{(a) The 1d ${\cal N}=(0,4)$ theory for the 5d $SU(N)$ with two anisymmetrics}
\label{SUwithAntisymmetrics}
\end{figure}

This theory is obtained from the $\mathbb{Z}_2$ orbifold without vector structure of the $Sp(N)+\asymm$ theory.
The corresponding 1d ${\cal N}=(0,4)$ gauge theory for $k$ D0-branes in this case is shown in Fig.~\ref{SUwithAntisymmetrics}.
There is an additional $U(1)_{\ell'}$ global symmetry in this case, since 
the symmetry rotating the two antisymmetrics 
is $U(2)_{\ell'} = SU(2)_{\ell'}\times U(1)_{\ell'}$. 
The $SU(2)_{\ell'}$ doublets, $Y$, $\chi_R$, and $\psi_R$, all carry one unit of charge
under $U(1)_{\ell'}$. All other mutiplets are neutral under this symmetry. 
The twisted hypermultiplet $Y$ transforms in the symmetric of $U(k)$, the Fermi multiplet $\chi_R$ transforms in
the antisymmetric, the hypermultiplet $q$ transforms in $({\bf k},\bar{\bf N})$, and the Fermi multiplet $\psi_R$ in 
$({\bf k},{\bf N})$. Note in particular the different representations of $q$ and $\psi_R$.
The 1d $U(k)$ theory admits a CS term inhereted from the 5d $SU(N)$ theory.
As before, this corresponds to a background $U(1)$ gauge charge $\kappa$.
In addition there can be a background $U(1)_{\ell'}$ global charge $\zeta$, corresponding to a CS term
for a background $U(1)_{\ell'}$ gauge field $\zeta \int A_{\ell'}$.\footnote{This comes from a mixed CS term in 5d coupling the 
$SU(N)$ gauge field to a background $U(1)_{\ell'}$ gauge field,
$\zeta \int A_{\ell'} \wedge \mbox{Tr}(F\wedge F)$. In the background of an instanton particle this reduces to $\zeta \int A_{\ell'}$.}

Let us now consider the reduced theory obtained by removing $q$ and $\psi_R$.
As in the ${\cal N}=2$ case in the previous section, the would-be Higgs branch of the reduced theory given by
$\varphi_H = Y_H = 0$ and $X_H = \mbox{diag}[x_1,\ldots x_k]$ is fully contained in the Coulomb branch, and therefore
does not include a remnant of the original Higgs branch.\footnote{In this case the would-be twisted Higgs branch is also contained in the Coulomb branch.}
Therefore the reduced theory should correctly account for the extra states without the additional correction factor. 
Indeed we do not expect such a correction since the orbifold without vector structure does not support fractional D0-branes.

However the CS parameter $\kappa$ and the background $U(1)_{\ell'}$ charge $\zeta$ of the reduced theory
are shifted from their values in the original theory:
\be
\label{CSshift}
\Delta \kappa = 2 \Delta \zeta = \left\{
\begin{array}{rl}
\pm 2 & N=2n+1 \\
0 & N=2n \, .
\end{array}
\right.
\ee
We can see this as follows.
For $N=2n$, the mass deformation that moves the D4-branes away from the D0-branes, and specifically away from the O8-plane,
breaks the 5d symmetry $SU(2n)$ to $SU(n)\times SU(n) \times U(1)$.
The fermions $\psi_R$ and $\psi_L$ decompose as 
$({\bf n},{\bf 1})_+ + ({\bf 1},{\bf n})_-$ and $({\bf \bar{n}},{\bf 1})_- + ({\bf 1},{\bf \bar{n}})_+$, respectively.
Since the fermion mass is proportional to the $U(1)$ charge and the spin, each one gives $N$ positive mass fermions and $N$
negative mass fermions.
The CS parameter gets contributions from both $\psi_R$ and $\psi_L$, and the background global charge $\zeta$ gets
contributions only from $\psi_R$. In either case the net shift vanishes since there are an equal number of negative and positive
contributions.
On the other hand for $N=2n+1$ the unbroken 5d symmetry is $SU(n)\times SU(n+1)\times U(1)$,
and the two fermions decompose as $({\bf n},{\bf 1})_+ + ({\bf 1},{\bf n +1})_-$
and $({\bf \bar{n}},{\bf 1})_- + ({\bf 1},{\bf \overline{n + 1}})_+$.
In this case there is a mismatch of positive and negative mass fermions which shifts $\kappa$ by
$\Delta\kappa = \pm 2\times 2 \times \frac{1}{2} = \pm 2$,
and shifts $\zeta$ by $\Delta\zeta = \pm 2 \times \frac{1}{2} = \pm 1$.
The common factor of 2 is due to the fact that $\psi_R$ and $\psi_L$ are both doublets (of $SU(2)_{\ell'}$ and $SU(2)_{r'}$, respectively). 
At this point it is difficult to fix the sign of the shift. 
However we will see below that the minus sign appears to be the correct one.


Our proposal for $Z_{extra}$ in this case is then given by
\be
\label{Proposal1}
Z_{extra}[\kappa,\zeta] = Z_{QM,red}[\kappa',\zeta'] \,,
\ee
where $\kappa' = \kappa +\Delta\kappa = ֿ\kappa \pm 2$ and $\zeta' = \zeta + \Delta\zeta = \zeta \pm 1$.
The one-instanton contribution to $Z_{QM,red}$ is given by
\begin{eqnarray}
\label{SU(0)InstantonIndex}
I_{1,red}[\kappa',\zeta'] &=& \frac{a^{\zeta'}}{2\pi i}\oint z^{\kappa'} 
I_1^A I_1^X I_1^Y \\ \nonumber 
&=& \mbox{} - Z_{\mathbb{C}^2\times \mathbb{C}^2/\mathbb{Z}_2} a ^{\zeta'-\frac{\kappa'}{2}} 
\left\{
\begin{array}{ll}
0 & \mbox{if} \;\; \kappa' = \mbox{odd} \\[5pt]
1+t^2 & \mbox{if} \;\; \kappa' = 0 \\[5pt]
t^{2-\frac{\kappa'}{2}}\chi_{\boldsymbol{\frac{\kappa'}{2}+1}}[v]-t^{-\frac{\kappa'}{2}}\chi_{\boldsymbol{\frac{\kappa'}{2}-1}}[v]& \mbox{if} \;\; \kappa' = \mbox{even} >0\\
t^{-\frac{\kappa'}{2}}\chi_{\boldsymbol{-\frac{\kappa'}{2}+1}}[v]-t^{2-\frac{\kappa'}{2}}\chi_{\boldsymbol{-\frac{\kappa'}{2}-1}}[v] & \mbox{if} \;\; \kappa' = \mbox{even} <0
\end{array}
\right.
\end{eqnarray}
where $\chi_{\textbf{d}}[v]$ is the character of the $d$-dimensional representation of $SU(2)_{\ell'}$, 
$a$ denotes the $U(1)_{\ell'}$ fugacity, and  we have defined
\be
\label{OrbifoldIndex}
Z_{\mathbb{C}^2\times \mathbb{C}^2/\mathbb{Z}_2} \equiv \frac{t^2}{(1-tu)(1-\frac{t}{u})(1-t^2v^2)(1- \frac{t^2}{v^2})} \,.
\ee
This factor is associated to the translational zero modes of a D0-brane 
moving in $\mathbb{C}^2\times \mathbb{C}^2/\mathbb{Z}_2$. 
The one-instanton contribution vanishes for any odd value of $\kappa'$
due to the non-vanishing background gauge charge.
For even values of $\kappa'$ the background charge can be cancelled by the $Y$ multiplet, resulting in a nonzero contribution.

We claim that the full multi-particle index of the reduced theory is given by
\be
\label{ZQMSU(0)}
Z_{QM,red} = \mbox{PE} [q I_{1,red}] \,,
\ee
namely that there are no D0-brane bound states.
As a test of this claim we have computed the two-instanton contribution for $\kappa'=0, \pm 2$: 
\be
I_{2,red} = \left\{
\begin{array}{ll}
a^{2\zeta'}\left(\chi_{\textbf{2}}[u] t^5+(1+\chi_{\textbf{3}}[u]+\chi_{\textbf{3}}[v])t^6+{\cal O}(t^7)\right) & \mbox{for} \;\; \kappa' = 0 \\[5pt]
a^{2\zeta' - 2}\left(t^6 + \chi_{\textbf{2}}[u] \chi_{\textbf{2}}[v]^2t^7+ {\cal O}(t^8)\right) & \mbox{for} \;\; \kappa' = \pm 2 \,.
\end{array}
\right.
\ee
This agrees with the $q^2$ term in the expansion of (\ref{ZQMSU(0)}) for $\kappa'=0,\pm 2$.

\subsection{Adding flavors}

\begin{figure}[H]
\center
\includegraphics[height=0.2\textwidth]{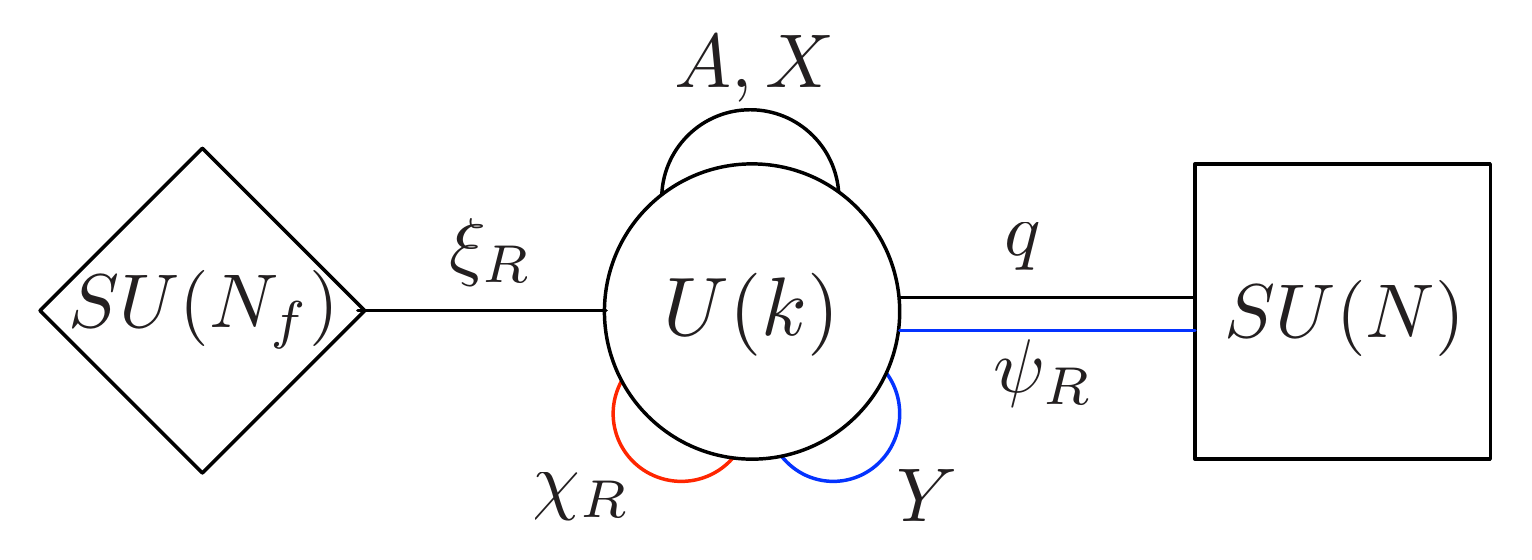} 
\caption{The 1d ${\cal N}=(0,4)$ theory for the 5d $SU(N)$ with two anisymmetrics and $N_f$ fundamentals.}
\label{SUwithAntisymmetricsFlavors}
\end{figure}

Adding $N_f$ fundamentals (flavors) to the 5d theory adds $N_f$ Fermi multiplets $\xi_R^{i}$ to the 1d ${\cal N}=(0,4)$ theory, Fig.~\ref{SUwithAntisymmetricsFlavors}. 
These contribute an additional factor of the form
\be
I_k^{\xi_R} = 
\prod_{I=1}^k \left[z_I^{N_f/2} \sum_{m=0}^{N_f} (-1)^m z_I^{-m} \chi_{[m]}[f]\right] \,,
\ee
where $\chi_{[m]}$ denotes the character of the antisymmetric $m$-tensor representation of $SU(N_f)$.
Our proposal for $Z_{extra}$ is the same as in (\ref{Proposal1}), namely $Z_{extra}[\kappa,\zeta] = Z_{QM,red}[\kappa',\zeta']$.
Here $\kappa$ (and therefore also $\kappa'$) takes values in $\mathbb{Z} + N_f/2$ due to the parity anomaly.
We expect that for a small number of flavors (\ref{ZQMSU(0)}) will continue to hold, namely that 
$Z_{QM,red} = \mbox{PE}\left[q I_{1,red}\right]$, 
where $I_{1,red}$ is the one-instanton contribution.
In the presence of $N_f$ flavors this is given by
\be
\label{redflavors}
I_{1,red}[\kappa',\zeta',N_f] &=& \frac{a^{\zeta'}}{2\pi i}\oint z^{\kappa'} 
I_1^A I_1^X I_1^Y I_1^{\xi_R} \nonumber \\
&=& \sum_{m=0}^{N_f}(-1)^{m} \, \chi_{[m]}[f] \, I_{1,red}[\kappa'+N_f/2-m,\zeta',N_f=0] \,,
\ee
where the $N_f=0$ one-instanton contribution on the r.h.s. is evaluated 
using (\ref{SU(0)InstantonIndex}) with $\kappa'$ replaced by $\kappa' + N_f/2 - m$.

\subsection{A consistency check}

For $SU(3)$ the antisymmetric representation is equivalent to the fundamental representation, or more precisely to the
antifundamental representation, so we can check our proposal in (\ref{Proposal1}) against the instanton partition function
for $SU(3)$ with two fundamentals.
For simplicity we set $\kappa=\zeta =0$.
Let us compare the two instanton partition functions at instanton number $k=1$.
Expanding to order $q$ we get
\be
Z_{inst} = \frac{Z_{QM}}{Z_{extra}} = 1 + \left(I_1 - I_{1,red}\right) q + {\cal O}(q^2) \,,
\ee
where $I_1$ is the $k=1$ contribution to $Z_{QM}$, and $I_{1,red}$ is the $k=1$ contribution to $Z_{QM,red}$.
For $SU(3) + 2\, \funda$ we have\footnote{For $SU(N)+N_f$ $k$-instantons there is an 
overall sign $(-1)^{k(\kappa+ N_f/2)}$ \cite{Bergman:2013aca}.}
\be
\label{SU3+2}
I_1^{SU(3)+2\,\funda}&=& \mbox{} - \frac{1}{2\pi i} \oint 
I_1^A I_1^X I_1^q  I_1^{\xi_R} \\
&=& \frac{f_1+f_2}{\sqrt{f_1f_2}}t^3 + \left( \frac{f_1+f_2}{\sqrt{f_1f_2}}\chi_{\textbf{2}}[u]- 
\sqrt{f_1f_2}\chi_{\textbf{3}}[s_i]+\frac{1}{\sqrt{f_1f_2}}\chi_{\Bar{\textbf{3}}}[s_i]\right)t^4+ {\cal O}(t^5) , \nonumber
\ee
where $f_{1,2}$ are the two flavor fugacities, and $s_i$ ($i=1,2,3$) are the gauge fugacities constrained by $\prod_i s_i = 1$.
The counterpart in the reduced QM vanishes in this case,
\be
I_{1,red}^{SU(3)+2\,\funda} = \mbox{} - \frac{1}{2\pi i}
 \oint 
I_1^A I_1^X  I_1^{\xi_R} = 0 \,,
\ee
since there are no contributing poles, as in the ${\cal N}=2$ $U(N)$ theory \cite{Hwang:2016gfw}.
For $SU(3) + 2\, \asymm$ we get
\be
I_1^{SU(3)+2\,\asymm} = \frac{1}{2\pi i} \oint I_1^A I_1^X I_1^q I_1^Y I_1^{\psi_R} =
\left(- {a}\chi_{\textbf{3}}[s_i]+\frac{1}{{a}}\chi_{\Bar{\textbf{3}}}[s_i]\right)t^4 + {\cal O}(t^5) \,,
\ee
and, using $\kappa' = 2\zeta' = \pm 2$,
\be
I_{1,red}^{SU(3)+2\,\asymm} = \mbox{} - t \chi_{\bf 2}[v]\, Z_{\mathbb{C}^2\times \mathbb{C}^2/\mathbb{Z}_2}
= \mbox{} - \chi_{\bf 2}[v] \left(t^3 + \chi_{\bf 2}[u]t^4 + {\cal O}(t^5)\right) \,.
\ee
Comparing with (\ref{SU3+2}) we see that the two instanton partition functions agree if we identify
$a=1/\sqrt{f_1 f_2}$ and $v = \sqrt{f_1/f_2}$.
Note that the agreement holds in this case for both choices of sign in $\Delta\kappa$ and $\Delta\zeta$. 
However for more general values of $\kappa$ and $\zeta$ the two computations agree only for the lower sign, namely
$\kappa' = \kappa - 2$ and $\zeta' = \zeta -1$.

We can further compare the two points of view with an additional flavor.
Take $\kappa = \frac{1}{2}$ and $\zeta = 0$.
For $SU(3) + 3\, \funda$ we get
\be 
I_1^{SU(3)+3\,\funda} &=& \mbox{} - \left(\frac{f_1 + f_2 + f_3}{\sqrt{f_1 f_2 f_3}}\right) t^3 \nonumber \\
&+& \frac{1}{\sqrt{f_1 f_2 f_3}}\Big(
(f_1 f_2+f_1 f_3+f_2 f_3)\chi_{\textbf{3}}[s_i]+\chi_{\Bar{\textbf{3}}}[s_i]-(f_1 + f_2 + f_3)\chi_{\textbf{2}}[u]\Big)t^4 \nonumber \\
&+& {\cal O}(t^5),
\ee
and as before $I_{1,red}^{SU(3)+3\,\funda} = 0$. 
On the other hand for $SU(3)+2\, \asymm +\funda$ we find
\be
I_1^{SU(3)+2\,\asymm + \funda} &=& \frac{1}{2\pi i} \oint z^{1/2} I_1^A I_1^X I_1^q I_1^Y I_1^{\psi_R}I_1^{\xi_R}  \nonumber \\
& = & -a \sqrt{f_3}t^3+\Big((\sqrt{f_3}\chi_{\textbf{2}}[v]+\frac{1}{a\sqrt{f_3}})\chi_{\textbf{3}}[s_i]+\frac{a}{\sqrt{f_3}}\chi_{\Bar{\textbf{3}}}[s_i] -a \sqrt{f_3}\chi_{\textbf{2}}[u]\Big)t^4 \nonumber \\
&+&{\cal O}(t^5)
\ee
where we denote by $f_3$ the fugacity of the single flavor,
and, with $\kappa' = -\frac{3}{2}$ and $\zeta' = -1$,
\be
I_{1,red}^{SU(3)+2\,\asymm + \funda}  = \mbox{}  \frac{t(v+\frac{1}{v})}{\sqrt{f_3}} \, Z_{\mathbb{C}^2\times \mathbb{C}^2/\mathbb{Z}_2} 
= \mbox{} \frac{\chi_{\bf 2}[v]}{\sqrt{f_3}} \left(t^3 + \chi_{\bf 2}[u] t^4 + {\cal O}(t^5)\right)\,.
\ee
%
The two instanton partition functions again agree once we identify $a=1/\sqrt{f_1 f_2}$ and $v = \sqrt{f_1/f_2}$.


\section{The $Sp(N_1)\times Sp(N_2) + (\funda,\funda)$ theory}

\begin{figure}[h]
\center
\includegraphics[width=0.3\textwidth]{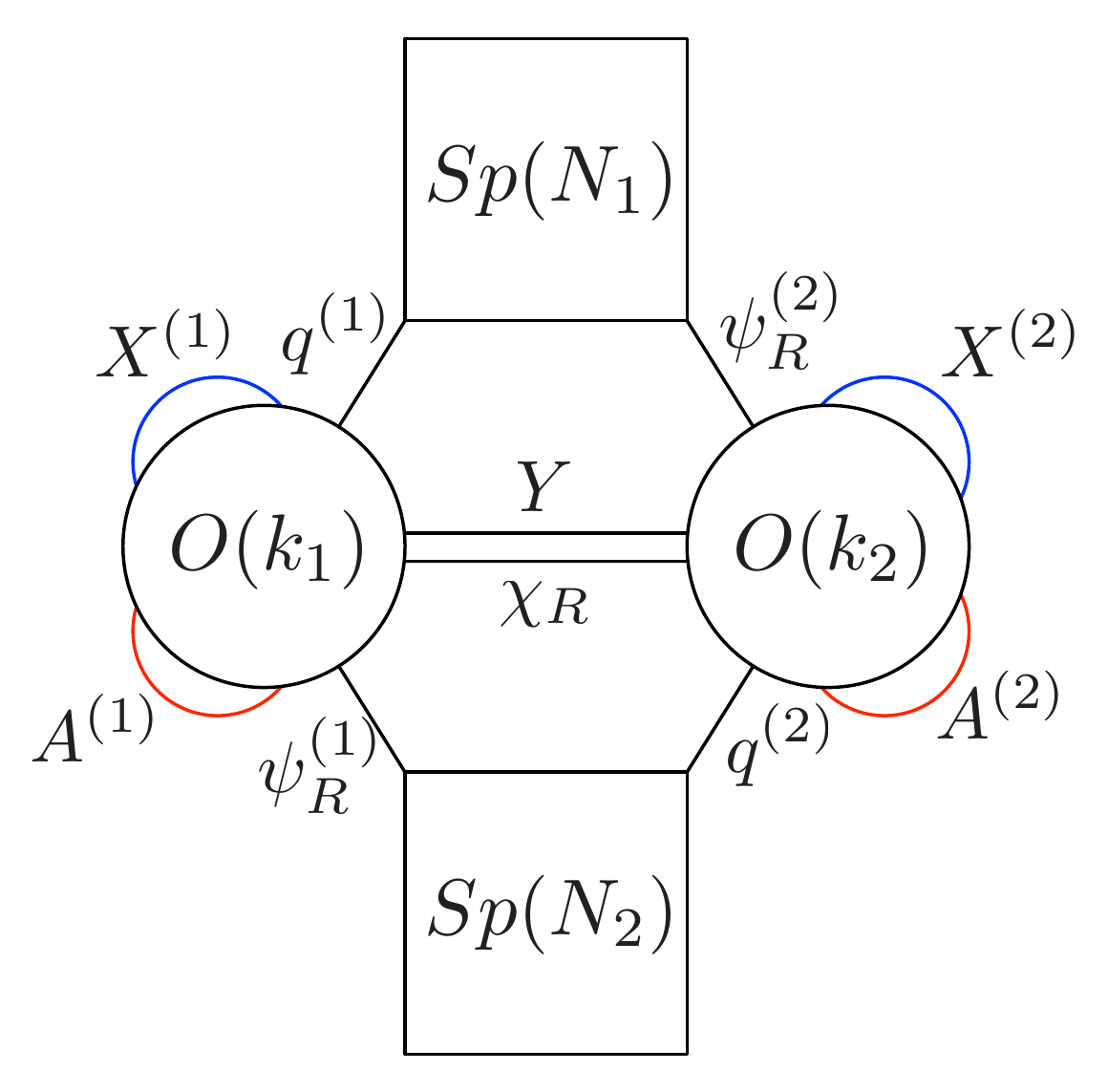} 
\caption{1d ${\cal N}=(0,4)$ theory for the 5d $Sp(N_1)\times Sp(N_2)$ with a bi-fundamental hypermultiplet. 
The \textcolor{blue}{blue} and \textcolor{red}{red} lines denote symmetric and antisymmetric representations, respectively.}
\label{SpSpwithBifundamental}
\end{figure}

This is the theory obtained from the $\mathbb{Z}_2$ orbifold with vector structure of the $Sp(N)+ \asymm$ theory.
Fig.~\ref{SpSpwithBifundamental} shows the structure of the corresponding 1d ${\cal N}=(0,4)$ gauge theory
for $k_1$ instantons of $Sp(N_1)$ and $k_2$ instantons of $Sp(N_2)$.
The 1d theory admits two discrete theta parameters $\theta_1$, $\theta_2$, that are inhereted from the 5d theory.
The reduced theory is obtained by integrating out the multiplets $q^{(i)}$ and $\psi_{R}^{(i)}$.
The Coulomb branch is given by two copies of the expressions in (\ref{CoulombBranchPhi}) and (\ref{CoulombBranchX}) for 
$\varphi_C^{(1,2)}$ and $X_C^{(1,2)}$ in the orthogonal group cases, and $Y_C=0$.
There is also a twisted Higgs branch on which $\varphi_{TH}^{(1,2)} = 0$ and  (we assume that $k_1\leq k_2$)
\be 
X_{TH}^{(1)} &=& \mbox{diag}[x_1,\ldots,x_{k_1}] \\
X_{TH}^{(2)} &=& \mbox{diag}[x_1,\ldots,x_{k_1},x_{k_1+1},\ldots,x_{k_2}]\\
Y_{TH} &=& \mbox{diag}[y_1,\ldots,y_{k_1}] \,.
\ee
On the other hand the Higgs branch of the reduced theory is given by $\varphi_H^{(1,2)} = Y_H = 0$ and 
$X_H^{(1,2)} = \mbox{diag}[x^{(1,2)}_1,\ldots ,x^{(1,2)}_{k_{1,2}}]$, where $x_a^{(1)}$ and $x_a^{(2)}$ are independent.
This is a separate branch, and is a remnant of the Higgs branch of the original theory.
It corresponds to fractional D0-branes confined to the orbifold fixed plane.
We therefore need to divide by the appropriate correction factor $\mbox{PE}[f_{D0_c}(q_1,q_2)]$.
We also need to take into account the shifted discrete theta parameters.
The fermion fields $\psi_L^{(i)}$ and $\psi_R^{(i)}$ will give $N_1+N_2$ positive mass fermions and $N_1 + N_2$ 
negative mass fermions charged under $O(k_i)$, and therefore $\theta_i' = \theta_i + \Delta\theta_i$, with
\beq
\label{theta shift}
\Delta \theta_i = \pi [(N_1 + N_2) \, \mbox{mod} \, 2].
\eeq

Our proposal in this case therefore takes the form
\be
Z_{extra}[\theta_1,\theta_2] = \frac{Z_{QM,red}[\theta_1',\theta_2']}{\mbox{PE}[f_{D0_c}(q_1,q_2)]} \,.
\ee
The multiparticle index of the reduced theory is given in general by
\be
Z_{QM,red} = 1 + I_{1,0}^{red} q_1 + I_{0,1}^{red} q_2 + I_{1,1}^{red} q_1 q_2 + I_{2,0}^{red} q_1^2
+ I_{0,2}^{red} q_2^2 +{\cal O}(q^3) \,,
\ee
where $I_{k_1,k_2}^{red}$ is the index of the $O(k_1)\times O(k_2)$ theory.
First, we would like to claim that this plethystically-exponentiates to
\be 
\label{ZQMSpSp}
Z_{QM,red} = \mbox{PE}[I_{1,0}^{red} \, q_1 + I_{0,1}^{red} \, q_2 + (I_{1,1}^{red} - I_{1,0}^{red} I_{0,1}^{red}) \, q_1 q_2] \,.
\ee
We will test this claim shortly by computing some higher order terms.
Bust first notice that the first and second terms in the plethystic exponent are the contributions of the fractional D0-branes, 
and the third term is the contribution of the bulk D0-brane.
The former are precisely cancelled by the denominator factor leaving
\be 
\label{ZextraSpSp}
Z_{extra} = \mbox{PE}[(I_{1,1}^{red} - I_{1,0}^{red} I_{0,1}^{red} ) \, q_1 q_2] \,.
\ee

Now let us compute.
The $(1,0)$  contribution is given by
\be
\label{I(1,0)Reduced}
I_{1,0}^{red} = \frac{1}{2}(1+e^{i\theta_1'}) \, I_1^X =  \frac{1}{2}(1+e^{i\theta_1'})  Z_{\mathbb{C}^2} \,, 
\ee
and the $(0,1)$ contribution $I_{0,1}^{red}$ is similarly given by replacing $\theta_1'$ by $\theta_2'$.
The $(1,1)$ contribution is given by
\be
I^{red}_{1,1} &=& \frac{1}{4}(I^X_{1})^2 \Big[
I_{1,1}^{Y++}I^{\chi++}_{1,1}
+ e^{i\theta'_1} I_{1,1}^{Y-+}I_{1,1}^{\chi-+} + e^{i\theta'_2} I_{1,1}^{Y+-}I_{1,1}^{\chi+-}
+ e^{i(\theta'_1+\theta'_2)} I_{1,1}^{Y--}I_{1,1}^{\chi--} \Big] \nonumber \\[5pt]
&=& \frac{t^3}{4(1-tu)^2(1- \frac{t}{u})^2}\left[
(1 + e^{i(\theta_1' + \theta_2')}) \frac{1- v^{\pm 1}u}{1- v^{\pm 1}t}
+ (e^{i\theta_1'} + e^{i\theta_2'}) \frac{1+ v^{\pm 1}u}{1+ v^{\pm 1}t}
\right] .
\ee
Putting these together, our claim in (\ref{ZQMSpSp}) becomes
\be
\label{PE Sp0^2}
Z_{QM,red} = \left\{
\begin{array}{ll}
\mbox{PE}[ (q_1 + q_2)Z_{\mathbb{C}^2} -(1+t^2)   q_1 q_2 Z_{\mathbb{C}^2\times \mathbb{C}^2/\mathbb{Z}_2}]
& (\theta'_1,\theta'_2) = (0,0)\\[5pt]
\mbox{PE}[-t  \chi_{\bf 2}[v]   q_1 q_2 Z_{\mathbb{C}^2\times \mathbb{C}^2/\mathbb{Z}_2} ] & (\theta'_1,\theta'_2) = (\pi,\pi) \\[5pt]
\mbox{PE}[ q_1 Z_{\mathbb{C}^2} ]
& (\theta'_1,\theta'_2) = (0,\pi) \\[5pt]
\mbox{PE}[q_2 Z_{\mathbb{C}^2} ]
& (\theta'_1,\theta'_2) = (\pi,0) \,,
\end{array}
\right.
\ee
where $Z_{\mathbb{C}^2\times \mathbb{C}^2/\mathbb{Z}_2}$ was defined in (\ref{OrbifoldIndex}), and 
\be
Z_{\mathbb{C}^2} \equiv \frac{t}{(1-tu)(1- \frac{t}{u})}\,.
\ee
To test this claim let us compute a couple of higher order terms.
The $(2,0)$ term should be given by 
\be
I^{red}_{2,0} &=& \frac{1}{2}\left[
\oint  I_{2}^{A+}I_{2}^{X+}
 + e^{i\theta_1'} I_{2}^{A-}I_{2}^{X-}\right] \nonumber \\
 &=&\left\{
 \begin{array}{ll}
\frac{ t^2 (1+t^2) }{(1+t u) (1+ \frac{t}{u})  (1-tu)^2 (1- \frac{t}{u})^2 }& (\theta'_1,\theta'_2) =(0,0),(0,\pi)\\[5pt]
0 & (\theta'_1,\theta'_2) = (\pi,0),(\pi,\pi) \,,
\end{array}
\right.
\ee
which agrees with the $q_1^2$ term in the expansion of (\ref{PE Sp0^2}).
A similar conclusion holds for the $(0,2)$ term.
The $(2,1)$ term should be given by
\be
I^{red}_{2,1} &=& \frac{1}{4}I_{1}^X\Big[
\oint I_{2}^{A+}I_{2}^{X+} (I_{2,1}^{Y++} I_{2,1}^{\chi_R++} + e^{i\theta'_2}I_{2,1}^{Y+-} I_{2,1}^{\chi_R+-}) \\\nonumber
&& \mbox{} +  e^{i\theta_1'} I_{2}^{A-}I_{2}^{X-}   
(I_{2,1}^{Y-+} I_{2,1}^{\chi_R-+} + e^{i\theta'_2}I_{2,1}^{Y--} I_{2,1}^{\chi_R--})
 \Big]\\\nonumber
 &=&\left\{
 \begin{array}{ll}
 \frac{(t+\frac{1}{t}) (v^2+\frac{1}{v^2}-u^2-\frac{1}{u^2})}{(t+\frac{1}{t}-u-\frac{1}{u})^3 (t+\frac{1}{t}+u+\frac{1}{u}) (v^2+\frac{1}{v^2}-t^2-\frac{1}{t^2})}& (\theta'_1,\theta'_2) =(0,0)\\[5pt]
0 & (\theta'_1,\theta'_2) = (\pi,0),(0,\pi),(\pi,\pi) \,,
 \end{array}
\right.
\ee
which agrees with the $q_1^2 q_2$ term.
A similar conclusion holds for the $(1,2)$ term.\\

Going back to $Z_{QM,red}$, we factor out the contribution of the fractional D0-branes given by $\mbox{PE}[I_{1,0}^{red} \, q_1 + I_{0,1}^{red} \, q_2]$, resulting with
\be
\label{SpSpClaim}
Z_{extra} = \left\{
\begin{array}{ll}
\mbox{PE}[-(1+t^2) q_1 q_2 Z_{\mathbb{C}^2\times \mathbb{C}^2/\mathbb{Z}_2} ] & (\theta'_1 ,\theta'_2) = (0,0) \\ [5pt]
\mbox{PE}[-t  \chi_{\bf 2}[v] q_1 q_2 Z_{\mathbb{C}^2\times \mathbb{C}^2/\mathbb{Z}_2} ] & (\theta'_1 ,\theta'_2) = (\pi,\pi)\\[5pt]
1 &(\theta'_1 ,\theta'_2)=(\pi,0),(0,\pi).
\end{array}
\right.
\ee


\subsection{A consistency check}

Since $Sp(1) \sim SU(2)$, we can check our proposal (\ref{SpSpClaim}) for the case of $Sp(1)\times Sp(1)$
against the instanton partition function obtained in the so-called ``$SU(N)$ formalism" by treating each $SU(2)$ as $U(2)/U(1)$ \cite{Bergman:2013aca} (see also appendix B).
Expanding to order $q^2$, our proposal gives
\be
Z_{inst} &=& \frac{Z_{QM}}{Z_{extra}}  \\
&=& 1 + I_{1,0} q_1 + I_{0,1} q_2 + I_{2,0} q_1^2 + I_{0,2} q_2^2
+\left(I_{1,1} - I_{1,1}^{red} + I_{1,0}^{red} I_{0,1}^{red}\right) q_1 q_2 + {\cal O}(q^3) \nonumber
\ee
The $(1,0)$ instanton term is given by
\be
I^{inst}_{1,0} &=& I_{1,0} \nonumber \\
&=& \frac{1}{2}\left[ I_1^{X+} I_1^{q+} I_1^{\psi_R+} + e^{i\theta_1} I_1^{X-} I_1^{q-} I_1^{\psi_R-}\right] \\\nonumber
&=& \frac{1}{2} \,
\frac{1}{t+\frac{1}{t}-u-\frac{1}{u}}\Bigg(\frac{v+\frac{1}{v}+s+\frac{1}{s}}{t+\frac{1}{t}+s+\frac{1}{s}}
+ e^{i\theta_1}\frac{v+\frac{1}{v}-s-\frac{1}{s}}{t+\frac{1}{t}-s-\frac{1}{s}}\Bigg) \,,
\ee
which agrees with the corresponding quantity computed in the $SU(N)$ formalism in \cite{Bergman:2013aca}.
The $(0,1)$ instanton term $I_{0,1}$ likewise agrees.
The $(2,0)$ instanton term is given by
\be
I^{inst}_{2,0} &= & I_{2,0} \nonumber \\
&=&\frac{1}{2}\Big(\oint I_2^{A+}I_2^{X+} I_2^{q+}I_2^{\psi_R+} + e^{i\theta_1} I_2^{A-} I_2^{X-}I_2^{q-}I_2^{\psi_R-}\Big) \\
&=& \left\{
\begin{array}{ll}
\chi_{\textbf{3}}[v] t^4+\chi_{\textbf{2}}[v](\chi_{\textbf{2}}[u]\chi_{\textbf{2}}[v]-\chi_{\textbf{2}}[s']\chi_{\textbf{2}}[s])t^5+\mathcal{O}(t^6) & \theta_1=0 \nonumber \\[5pt]
\chi_{\textbf{3}}[s'] t^4+\chi_{\textbf{2}}[s'](\chi_{\textbf{2}}[u]\chi_{\textbf{2}}[s']-\chi_{\textbf{2}}[v]\chi_{\textbf{2}}[s])t^5+\mathcal{O}(t^6) & \theta_1=\pi \nonumber
\end{array}
\right.
\ee
where $s,s'$ are the fugacities associated to the first and second $Sp(1)$ factors, respectively.
This also agrees with the corresponding quantity computed in the $SU(N)$ formalism in \cite{Bergman:2013aca}.
The $(0,2)$ instanton term $I_{0,2}$ likewise agrees.
The first term where the correction factor (\ref{SpSpClaim}) is relevant is the $(1,1)$ instanton term, given by
\be
I^{inst}_{1,1} &=& I_{1,1} - I_{1,1}^{red} + I_{1,0}^{red} I_{0,1}^{red} \nonumber \\
&=&  I_{1,1} - Z_{\mathbb{C}^2\times \mathbb{C}^2/\mathbb{Z}_2}\times
\left\{
\begin{array}{ll}
-(1+t^2) & (\theta_1,\theta_2) = (0,0)\\[5pt]
-t \chi_{\bf 2}[v]  & (\theta_1,\theta_2) = (\pi,\pi) \\[5pt]
0 & (\theta_1,\theta_2) = (\pi,0),(0,\pi)
\end{array}
\right.
\ee
where
\be
I_{1,1} &=&
\frac{1}{4}[I_1^X]^2
\Big(I_{1,1}^{Y++}I_{1,1}^{\chi_R++}[I_1^{q+}]^2 [I_1^{\psi_R+}]^2
+e^{i\theta_1} I_{1,1}^{Y+-}I_{1,1}^{\chi_R+-} I_1^{q+} I_1^{q-} I_1^{\psi_R+} I_1^{\psi_R-} \\  \nonumber
&+&
e^{i\theta_2} I_{1,1}^{Y-+}I_{1,1}^{\chi_R-+} I_1^{q+} I_1^{q-} I_1^{\psi_R+} I_1^{\psi_R-} 
+e^{i(\theta_1+\theta_2)} I_{1,1}^{Y--}I_{1,1}^{\chi_R--}[I_1^{q-}]^2 [I_1^{\psi_R-}]^2 \Big) 
\ee
resulting with 
\be
I^{inst}_{1,1} =
\left\{
\begin{array}{ll}
t^2 + \chi_{\bf{2}}[u] t^3 + 
\chi_{\bf{3}}[u] \chi_{\bf{3}}[v]t^4 + & \\
\quad \Big(\chi_{\bf{2}}[u](\chi_{\bf{3}}[u]+2\chi_{\bf{3}}[v]) -\chi_{\bf{2}}[s]\chi_{\bf{2}}[s']\chi_{\bf{2}}[v]\Big)t^5+\mathcal{O}(t^6)
& (\theta_1,\theta_2)=  (0,0) \\[10pt]
 \chi_{\bf{2}}[v] t^3 + 
\chi_{\bf{2}}[u] \chi_{\bf{2}}[v]t^4+ & \nonumber \\
\quad \Big(
\chi_{\bf{2}}[s]\chi_{\bf{2}}[s']\chi_{\bf{2}}[u]+(\chi_{\bf{3}}[u]-1)\chi_{\bf{2}}[v]\Big)t^5+\mathcal{O}(t^6)
& (\theta_1,\theta_2) = (\pi,\pi) \\[10pt]
\chi_{\bf{2}}[v]\Big(\chi_{\bf{2}}[u]\chi_{\bf{2}}[s']+\chi_{\bf{2}}[s]\chi_{\bf{2}}[v]\Big)t^5+\mathcal{O}(t^6)
& (\theta_1,\theta_2)=  (0,\pi) \\[10pt]
\chi_{\bf{2}}[v]\Big(\chi_{\bf{2}}[u]\chi_{\bf{2}}[s]+\chi_{\bf{2}}[s']\chi_{\bf{2}}[v]\Big)t^5+\mathcal{O}(t^6)
& (\theta_1,\theta_2)=  (\pi,0)
\end{array}
\right.
\ee
which agrees with the $SU(N)$ formalism computation in \cite{Bergman:2013aca}.
Note that in this case $\theta_{1,2}' = \theta_{1,2}$.


\subsection{Another consistency check}

As one more consistency check we will consider the $(1,1)$ instanton sector of the $Sp(1)\times Sp(2)$ theory,
where we can compare our result to the result obtained by working in the $SU(N)$ formalism for the $Sp(1)=SU(2)$ factor.
This is the simplest instance where both aspects of our proposal are relevant.
One has to account for the shift in the discrete theta parameters, as well as for the contribution of the fractional D0-branes.
The resulting $(1,1)$ instanton term is given by
\be
\label{Sp1Sp2}
I^{inst}_{1,1} &=& I_{1,1}[\theta_1,\theta_2] - I_{1,1}^{red}[\theta_1+\pi,\theta_2+\pi] + I_{1,0}^{red}[\theta_1+\pi,\theta_2+\pi]
 I_{0,1}^{red}[\theta_1+\pi,\theta_2+\pi] \nonumber \\[10pt]
&=& \left\{
\begin{array}{ll}
\chi_{\bf{2}}[v]t^3 +\chi_{\bf{2}}[v] \chi_{\bf{2}}[u] t^4 + & \\
\quad \chi_{\bf{2}}[v](\chi_{\bf{3}}[v]+\chi_{\bf{3}}[u]-1)t^5 +\mathcal{O}(t^6) & (\theta_1,\theta_2)=  (0,0) \\[10pt]
t^2+\chi_{\bf{2}}[u] t^3+ (\chi_{\bf{3}}[u]+\chi_{\bf{3}}[v])t^4+ & \\
\quad \chi_{\bf{2}}[u](\chi_{\bf{3}}[u]+\chi_{\bf{3}}[v]-1)t^5 
+\mathcal{O}(t^6) & (\theta_1,\theta_2) = (\pi,\pi) \\[10pt]
\mathcal{O}(t^6) & (\theta_1,\theta_2)=  (0,\pi), (\pi,0)
\end{array}
\right.
\ee
This agrees with the result of the $SU(N)$ formalism
 (see appendix B).

\subsection{Adding flavors} 

\begin{figure}[h]
\center
\includegraphics[width=0.5\textwidth]{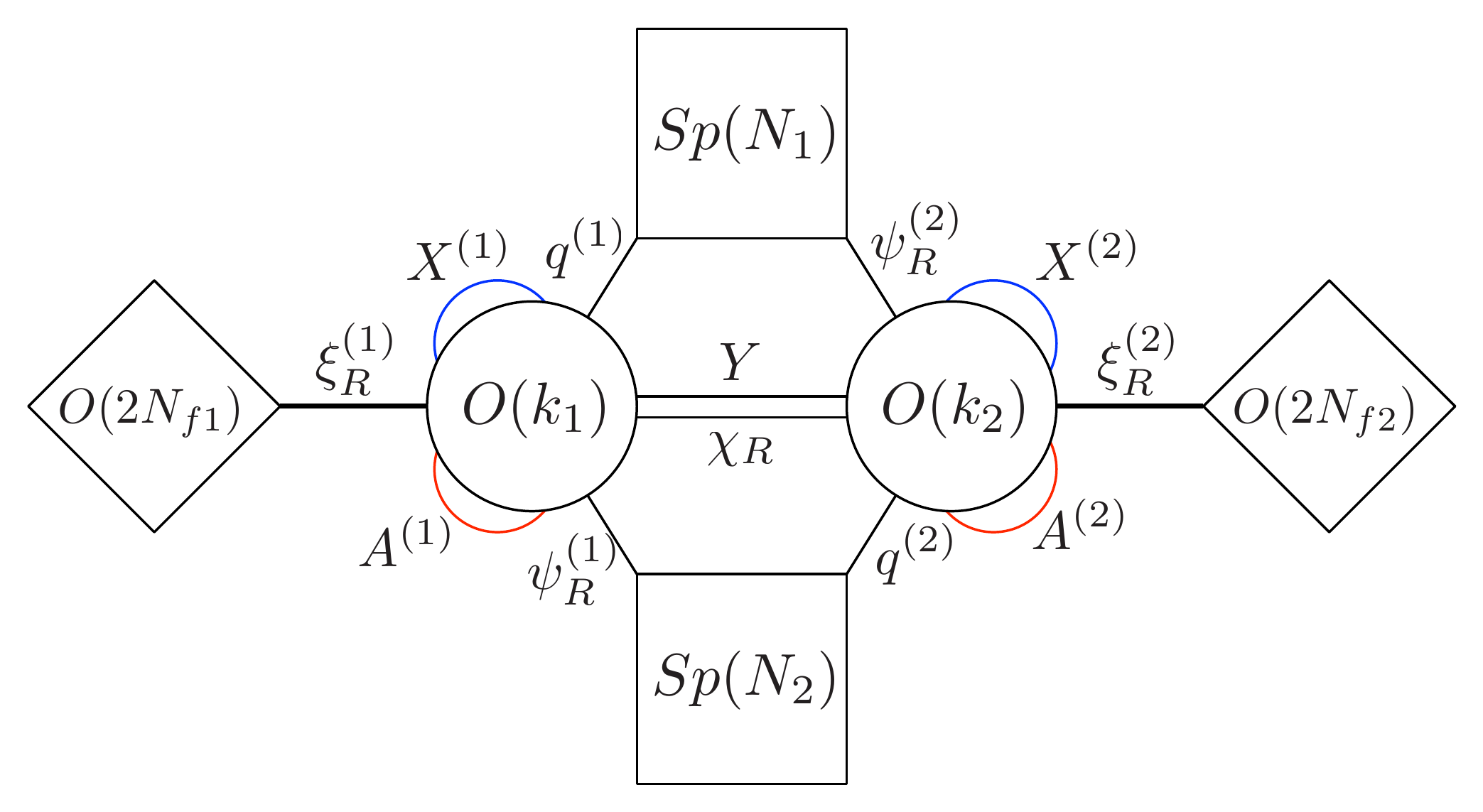} 
\caption{1d ${\cal N}=(0,4)$ theory for the 5d $Sp(N_1)\times Sp(N_2)$ with a bi-fundamental hypermultiplet
and fundamental hypermultiplets.}
\label{SpSpwithFlavors}
\end{figure}

Adding flavors to the 5d gauge theory corresponds to adding Fermi multiplets $\xi_R$ to the 1d gauge theory as shown in Fig.~\ref{SpSpwithFlavors}.
These endow the states with spinor charges under the global $O(2{N_f}_1)$ and $O(2{N_f}_2)$ symmetries.
There are two Weyl spinors that we denote ${\bf S}$ and ${\bf S}'$.
Since under the parity element of the $O(k)$ gauge symmetry ${\bf S}$ is even and ${\bf S}'$ is odd,
there is an additional sign in the projection in the latter case.
For example, the expression for $I_{1,0}^{red}$ in (\ref{I(1,0)Reduced}) is replaced with
\be
I_{1,0}^{red} = \frac{1}{2}  \left[(1+e^{i\theta_1'}) \chi_{\bf S_1}[f_1] + (1- e^{i\theta_1'})\chi_{\bf S_1'}[f_1]\right] Z_{\mathbb{C}^2} \,.
\ee
As before, we expect that for a small number of flavors equation (\ref{ZQMSpSp}) continues to hold. In  this case
\beq
Z_{QM}^{red}=\left\{
\begin{array}{ll}
\mbox{PE}\big[-\left((1+t^2)\chi_{\bf{S_1}}\chi_{\bf{S_2}}+
t \chi_{\bf 2}[v] \chi_{\bf{S_1'}}\chi_{\bf{S_2'}}\right) q_1 q_2 Z_{\mathbb{C}^2\times \mathbb{C}^2/\mathbb{Z}_2} & \\[5pt]
 \qquad\qquad \mbox{} +  \left(\chi_{\bf S_1} q_1 + \chi_{\bf S_2} q_2 \right) Z_{\mathbb{C}^2}\big] &  (\theta_1',\theta_2')=(0,0)\\[10pt]
\mbox{PE}\big[-\left(t \chi_{\bf 2}[v] \chi_{\bf{S_1}}\chi_{\bf{S_2}}+
(1+t^2)\chi_{\bf{S_1'}}\chi_{\bf{S_2'}}\right)q_1q_2 Z_{\mathbb{C}^2\times \mathbb{C}^2/\mathbb{Z}_2} & \\[5pt]
\qquad\qquad \mbox{} +  \left(\chi_{\bf S_1'} q_1 + \chi_{\bf S_2'} q_2 \right) Z_{\mathbb{C}^2} \big]
&  (\theta_1',\theta_2')=(\pi,\pi) \\[10pt]
\mbox{PE}\left[(\chi_{\bf S_1} q_1 + \chi_{\bf S_2'} q_2)Z_{\mathbb{C}^2}\right] & (\theta_1',\theta_2') = (0,\pi) \\[10pt]
\mbox{PE}\left[(\chi_{\bf S_1'} q_1 + \chi_{\bf S_2} q_2)Z_{\mathbb{C}^2}\right] & (\theta_1',\theta_2') = (\pi,0) 
\end{array}
\right.
\eeq
where for brevity we have suppressed the flavor fugacity arguments of the spinor characters.
Factoring out the contribution of the fractional D0-branes as before we get
\be
Z_{extra} = \left\{
\begin{array}{ll}
\mbox{PE}\left[-\left((1+t^2)\chi_{\bf{S_1}}\chi_{\bf{S_2}}+
t \chi_{\bf 2}[v] \chi_{\bf{S_1'}}\chi_{\bf{S_2'}}\right)q_1q_2 Z_{\mathbb{C}^2\times \mathbb{C}^2/\mathbb{Z}_2}\right] &  (\theta_1',\theta_2')=(0,0)\\[10pt]
\mbox{PE}\left[-\left(t \chi_{\bf 2}[v] \chi_{\bf{S_1}}\chi_{\bf{S_2}}+
(1+t^2)\chi_{\bf{S_1'}}\chi_{\bf{S_2'}}\right)q_1q_2 Z_{\mathbb{C}^2\times \mathbb{C}^2/\mathbb{Z}_2}\right]& (\theta_1',\theta_2')=(\pi,\pi) \\[10pt]
1& (\theta_1',\theta_2') = (0,\pi) \\[10pt]
1 & (\theta_1',\theta_2') = (\pi,0) .
\end{array}
\right.
\ee

\section{Conclusions}

In this note we have refined the procedure suggested in \cite{Kim:2012gu,Hwang:2014uwa,Hwang:2016gfw} for obtaining the  5d Nekarsov partition function 
using the 1d ${\cal N}=(0,4)$ gauge theory constructed from the ADHM data.
Our proposal correctly removes the contribution of the extra branches that are generically present when the 5d theory has matter in higher 
representations of the gauge symmetry, and are not associated to the instanton moduli space.
In particular it accounts for the possible overcounting in the reduced 1d gauge theory in cases where it maintains a remnant of the original Higgs branch,
and for the possible shifts in the parameters of the reduced 1d theory relative to the original 1d theory.

Our proposal is consistent with the results of \cite{Hwang:2016gfw} for the ${\cal N}=2$ theories and with the results of \cite{Kim:2012gu,Hwang:2014uwa} for the ${\cal N}=1$
$Sp(N) + \asymm$ theory.
We have applied it also to the theories obtained as $\mathbb{Z}_2$ orbifolds of the $Sp(N) + \asymm$ theory:
the $SU(N) + 2\, \asymm$ theory and the $Sp(N_1)\times Sp(N_2) +(\funda,\funda)$ theory,
where no previous results exist, and tested our proposal by 
exploiting simple isomorphisms between low-rank Lie groups. 

An important application of this work is to test the dualities proposed in \cite{Bergman:2013aca,Zafrir:2014ywa} that involve these theories
by comparing the superconformal indices.
So far this has only been done in the fractional instanton sectors, where the refinements described in this paper are absent.
Our proposal makes it possible to extend the computation to include any $(k_1,k_2)$ instanton.


\section*{Acknowledgements}
We thank Gabi Zafrir for many useful discussions and participation in early stages of this project.
This work is supported in part by the Israel Science Foundation under grant no. 352/13,
and by the US-Israel Binational Science Foundation under grant no. 2012-041.


\begin{appendices}

\section{The ${\cal N}=(0,4)$ index}



The $(0,2)$ QM index relevant for the ${\cal N}=(0,4)$ theories that we discuss is defined as
\be
I_k^{QM}=\text{Tr}_k[(-1)^Fe^{-\beta\{Q,Q^\dag\}}e^{-2\epsilon_+(J_r+J_{r'})}e^{-2\epsilon_-J_{\ell}}
e^{-2mJ_{\ell'}}e^{-\alpha_i\Pi_i} e^{-w_a F_a}] \,,
\eeq
where the $J$'s are the generators of the Cartan subgroups of the corresponding $SU(2)$ symmetries,
$\Pi_i$ are the generators of the Cartan subgroup of $G_{D4}$, 
and $F_a$ 
are the generators of the Cartan subgroup of $G_{D8}$.
The QM Euclidean action is compactified on a circle with circumference $\beta$. The path integral then reduces in the free field limit to Gaussian integrals around zero modes of the gauge field $A_0$ and the vector multiplet scalar $\varphi$. 
The two are combined into a complexified holonomies $\phi=\beta (iA_0+\varphi)$
taking value in the maximal torus of the complexified gauge group $G_{D0}$,
\beq
e^\phi = \left\{
\begin{array}{ll}
\text{diag}(e^{\phi_1},...,e^{\phi_k})&\in  U(k)\\
\text{diag}(e^{\sigma_2\phi_1},...,e^{\sigma_2\phi_n})&\in  O(2n)_{+}\\
\text{diag}(e^{\sigma_2\phi_1},...,e^{\sigma_2\phi_{n-1}},\sigma_3)&\in  O(2n)_{-}\\
\text{diag}(e^{\sigma_2\phi_1},...,e^{\sigma_2\phi_n},1)&\in  O(2n+1)_{+}\\
\text{diag}(e^{\sigma_2\phi_1},...,e^{\sigma_2\phi_n},-1)&\in  O(2n+1)_{-}\\
\text{diag}(e^{\sigma_3\phi_1},...,e^{\sigma_3\phi_k})&\in  Sp(k) \,.
\end{array}
\right.
\eeq
The result can in general be expressed as a contour integral in $\phi$ of a product of contributions of the different ${\cal N}=(0,4)$ multiplets:
\begin{equation}
\label{QMIndex0}
I_k^{QM} =  \oint_{z_I} I_k^A(z_I,t) I_k^X(z_I,t,u)  I_k^q(z_I,t,s_i) I_k^Y(z_I,t,v) I_k^{\chi_R}(z_I,u,v) I_k^{\psi_R}(z_I,v,s_i) I_k^{\xi}(z_I,f_a) ,
\end{equation}
where following \cite{Hwang:2014uwa,Hwang:2016gfw} we have defined the fugacities
\be
t\equiv e^{-\epsilon_+} \;, \; u\equiv e^{-\epsilon_-} \;, \; v\equiv e^{-m}  \; , \; s_i \equiv e^{\alpha_i} \; , \; f_a \equiv e^{-w_a} \; , \;
z_I \equiv e^{-\phi_I} \,.
\ee
The contributions of the different $(0,4)$ multiplets can be determined by decomposing them into $(0,2)$ multiplets,
whose contributions have a well known structure \cite{Hwang:2014uwa,Hwang:2016gfw}.
We will give the explicit expressions for the different theories below.
In particular the contribution of the vector multiplet includes the Haar measure $[dz_I]$.

Preforming the integral one must provide a prescription for the contours, namely for how to deal with the poles.
This question was answered for 2d gauge theories with ${\cal N}=(2,2)$ and ${\cal N}=(0,2)$ supersymmetry in 
\cite{Benini:2013nda,Benini:2013xpa} using the Jeffrey-Kirwan (JK) residue rule. 
The quantity of interest in that case is the elliptic genus, a refinement of the partition function on the torus.
Since ${\cal N}=(0,4)$ QM may be regarded as the dimensional reduction of a 2d ${\cal N}=(0,4)$ field theory, 
which is a special case of a 2d ${\cal N}=(0,2)$ theory, the same rule can be applied to the QM index,
which corresponds to a limit of the 2d elliptic genus. 
This is the approach taken in  \cite{Hwang:2014uwa}.
The authors of \cite{Hwang:2014uwa} have also noted that the JK prescription is equivalent, at least in the examples they studied,
to integrating each of the complex $U(1)$ holonomies separately on a unit circle, and applying the following residue rules:
poles arising from the 5d gauge multiplet contribute when they are inside the contour,
poles arising from 5d matter hypermultiplets contribute when they are outside the contour, and poles at the origin do not contribute.
The final result is 
\beq
I_k^{QM}=\frac{1}{|W_{G_{D0}}|}\sum\text{JK-Res}
\eeq
where $W_{G_{D0}}$ is the Weyl group of $G_{D0}$. In particular
\beq
&&|W|_{U(k)}=k!,\quad |W|_{O(2n)_+}=2^{n-1}n!,\quad |W|_{O(2n)_-}=2^{n-1}(n-1)!\\\nonumber
&&|W|_{O(2n+1)_+}=2^{n}n!,\quad |W|_{O(2n+1)_-}=2^n n!, \quad |W|_{Sp(k)}=k!.
\eeq
For $G_{D0}=O$ one must sum over the two holonomy sectors $O_+$ and $O_-$, with a relative weight given by the discrete theta
parameter as follows \cite{Bergman:2013ala}:
\beq
I^{QM}_k=\begin{cases}
\begin{array}{ll}
\frac{1}{2}( I^k_+ + I_-^k),  & \theta =0 \\[5pt]
\frac{(-1)^k}{2}( I^k_+-I_-^k) & \theta =\pi
\end{array}
\end{cases}
\eeq

In the remaining part of this appendix we will present the formulas for the contributions of the $(0,4)$ multiplets that arise for the 5d ${\cal N}=1$ theories
featured in this paper.
This information can also be extracted from \cite{Bergman:2013aca}, where it is organized by the 5d multiplets. 
For brevity we will adopt the shorthand $1 - u^{\pm 1} t = (1-ut)(1-u^{-1}t)$, etc.

\subsection{$Sp(N) + \asymm$}

The content of the 1d theory in this case is shown in Fig.~\ref{SpNwithAntisymmetric}.
There are four cases to consider, $O(k)_{\epsilon}$ with $k=2n, 2n+1$ and $\epsilon= \pm$.
For $k=2n+1$ we have:
\be\nonumber
I_k^{A,\epsilon} &=& \frac{\epsilon^n}{2^n n!} \Big(\frac{1}{t}-t\Big)^n 
\prod_{I=1}^n \frac{(1- \epsilon z_I^{\pm 1} t^2)(z_I + \frac{1}{z_I} - 2\epsilon)dz_I}{t^{2}z_I} 
\prod_{I<J}^n \frac{(1- z_I^{\pm 1} z_J^{\pm 1} t^2)(z_I + \frac{1}{z_I} - z_J - \frac{1}{z_J})^2}{t^{4}}\\
\\
I_k^{X,\epsilon} &=& 
\frac{t^{n+1}}{(1-u^{\pm 1}t)^{n+1}}
\prod_{I=1}^n \frac{t^4}{(1 - \epsilon z_I^{\pm 1} u^{\pm 1} t)(1-z_I^{\pm 2} u^{\pm 1}t)}
 \prod_{I < J}^n
 \frac{t^4}{(1-z_I^{\pm 1} z_J^{\pm 1} u^{\pm 1}t)} \\[10pt]
I_k^{q,\epsilon} &=& 
\prod_{i=1}^N \frac{t}{(1- \epsilon s_i^{\pm 1} t)}
\prod_{I=1}^n \prod_{i=1}^N \frac{t^2}{(1- s_i^{\pm 1} z_I^{\pm 1} t)}\\[10pt]
I_k^{Y,\epsilon} &=&
\frac{t^{n+1}}{(1-v^{\pm 1}t)^{n+1}}
\prod_{I=1}^n \frac{t^4}{(1 - \epsilon z_I^{\pm 1} v^{\pm 1} t)(1-z_I^{\pm 2} v^{\pm 1}t)}
 \prod_{I < J}^n
 \frac{t^4}{(1-z_I^{\pm 1} z_J^{\pm 1} v^{\pm 1}t)} \\[10pt]
I_k^{\psi_R,\epsilon}&=& \prod_{i=1}^N \frac{(1- \epsilon s_i^{\pm 1}v)}{v}
\prod_{I=1}^n \prod_{i=1}^N \frac{(1 - s_i^{\pm 1} z_I^{\pm 1}v)}{v^2}\\[10pt]
I_k^{\chi_R,\epsilon} &=& \frac{(1-v^{\pm 1}u)^n}{u^n}
\prod_{I=1}^n \frac{(1- \epsilon z_I^{\pm 1} v^{\pm 1} u)}{u^2}
\prod_{I<J}^n \frac{(1-z_I^{\pm 1} z_J^{\pm 1} v^{\pm 1}u)}{u^4}\\[10pt]
I_k^{\xi_R,\epsilon} &=& \prod_{a=1}^{N_f} \prod_{I=1}^n (f_a + \frac{1}{f_a} - z_I - \frac{1}{z_I}) 
\prod_{a=1}^{N_f} (\sqrt{f_a} - \epsilon \frac{1}{\sqrt{f_a}})
\ee
For $k=2n$ we have:
\be
I_k^{A,\epsilon} &=& \left\{
\begin{array}{ll}
 \frac{\Big(\frac{1}{t}-t\Big)^n }{2^{n-1}n!}
{\displaystyle\prod_{I=1}^n} \frac{dz_I}{z_I} 
{\displaystyle\prod_{I<J}^n} \frac{(1- z_I^{\pm 1} z_J^{\pm 1} t^2)(z_I + \frac{1}{z_I} - z_J - \frac{1}{z_J})^2}{t^{4}} & \epsilon = + \\[10pt]
\frac{(-1)^n\Big(\frac{1}{t}-t\Big)^{n-1}\Big(\frac{1}{t}+t\Big) }{2^{n-1} (n-1)!} 
{\displaystyle\prod_{I=1}^{n-1}}\frac{(1- z_I^{\pm 2} t^4)(z_I^2 + \frac{1}{z_I^2} - 2) dz_I}{t^4 z_I}
{\displaystyle\prod_{I<J}^{n-1}} \frac{(1- z_I^{\pm 1} z_J^{\pm 1} t^2)(z_I + \frac{1}{z_I} - z_J - \frac{1}{z_J})^2}{t^{4}} & \epsilon = -
\end{array}
\right.\\[10pt]
I_k^{X,\epsilon} &=& \left\{
\begin{array}{ll}
\frac{t^n}{(1-u^{\pm 1}t)^n}
{\displaystyle\prod_{I=1}^n}
\frac{t^2}{(1-z_I^{\pm 2} u^{\pm 1}t) }
{\displaystyle \prod_{I< J}^n}
 \frac{t^4}{(1-z_I^{\pm 1} z_J^{\pm 1} u^{\pm 1}t)} & \epsilon = + \\ [10pt]
\frac{t^{n+2}}{(1+ u^{\pm 1}t)(1-u^{\pm 1}t)^{n+1}}
{\displaystyle\prod_{I=1}^{n-1}}
\frac{t^4}{(1-z_I^{\pm 2}u^{\pm 1}t)
(1-z_I^{\pm 2} u^{\pm 1}t) }
{\displaystyle \prod_{I< J}^{n-1}}
 \frac{t^4}{(1-z_I^{\pm 1} z_J^{\pm 1} u^{\pm 1}t)} & \epsilon = -
 \end{array}
 \right.\\[10pt]
I_k^{q,\epsilon} &=& \left\{
\begin{array}{ll}
{\displaystyle\prod_{I=1}^n} 
{\displaystyle\prod_{i=1}^N} \frac{t^2}{(1- s_i^{\pm 1} z_I^{\pm 1} t)} & \epsilon = +\\
{\displaystyle\prod_{i=1}^N} \frac{t^2}{(1-s_i^{\pm 2} t^2)}
{\displaystyle\prod_{I=1}^{n-1}} 
{\displaystyle\prod_{i=1}^N} \frac{t^2}{(1- s_i^{\pm 1} z_I^{\pm 1} t)} & \epsilon = -
\end{array}
\right.\\[10pt]
I_k^{Y,\epsilon} &=& \left\{
\begin{array}{ll}
\frac{t^n}{(1-v^{\pm 1}t)^n}
{\displaystyle\prod_{I=1}^n}
\frac{t^2}{(1-z_I^{\pm 2} v^{\pm 1}t) }
{\displaystyle \prod_{I< J}^n}
 \frac{t^4}{(1-z_I^{\pm 1} z_J^{\pm 1} v^{\pm 1}t)} & \epsilon = + \\ [10pt]
 \frac{t^{n+2}}{(1+ v^{\pm 1}t)(1-v^{\pm 1}t)^{n+1}}
{\displaystyle\prod_{I=1}^{n-1}}
\frac{t^4}{(1-z_I^{\pm 2}v^{\pm 1}t)
(1-z_I^{\pm 2} v^{\pm 1}t) }
{\displaystyle \prod_{I< J}^{n-1}}
 \frac{t^4}{(1-z_I^{\pm 1} z_J^{\pm 1} v^{\pm 1}t)} & \epsilon = -
 \end{array}
 \right.\\[10pt]
I_k^{\psi_R,\epsilon} &=& \left\{
\begin{array}{ll}
{\displaystyle\prod_{I=1}^n}
{\displaystyle \prod_{i=1}^N} \frac{(1 - s_i^{\pm 1} z_I^{\pm 1}v)}{v^2} & \epsilon=+\\[10pt]
{\displaystyle\prod_{i=1}^N} \frac{(1-s_i^{\pm 2}v^2)}{v^2}
{\displaystyle\prod_{I=1}^{n-1}} 
{\displaystyle\prod_{i=1}^N} \frac{(1 - s_i^{\pm 1} z_I^{\pm 1}v)}{v^2} & \epsilon=-
\end{array}
\right.\\[10pt]
I_k^{\chi_R,\epsilon} &=& \left\{
\begin{array}{ll}
\frac{(1-v^{\pm 1}u)^n}{u^n}
{\displaystyle\prod_{I<J}^n} \frac{(1-z_I^{\pm 1} z_J^{\pm 1} v^{\pm 1}u)}{u^4} & \epsilon=+\\[10pt]
\frac{(1+v^{\pm 1}u)(1-v^{\pm 1}u)^{n-1}}{u^n}
{\displaystyle\prod_{I=1}^{n-1}} \frac{(1- z_I^{\pm 2} v^{\pm 2}u^2)}{u^4}
{\displaystyle\prod_{I<J}^{n-1}}  \frac{(1-z_I^{\pm 1} z_J^{\pm 1} v^{\pm 1}u)}{u^4}  & \epsilon = -
\end{array}
\right.\\[10pt]
I_k^{\xi_R,\epsilon} &=& \left\{
\begin{array}{ll}
{\displaystyle\prod_{a=1}^{N_f}} 
{\displaystyle\prod_{I=1}^n} (f_a + \frac{1}{f_a} - z_I - \frac{1}{z_I}) & \epsilon = +\\[10pt]
{\displaystyle\prod_{a=1}^{N_f}} 
{\displaystyle\prod_{I=1}^{n-1}} (f_a + \frac{1}{f_a} - z_I - \frac{1}{z_I}) 
{\displaystyle\prod_{a=1}^{N_f}} (f_a - \frac{1}{f_a}) & \epsilon = -
\end{array}
\right.
\ee

\subsection{$SU(N)+ 2\, \asymm$}

The content of the 1d theory in this case is shown in Fig.~\ref{SUwithAntisymmetricsFlavors}.
The contributions of the different 1d $(0,4)$ multiplets are given by:
\be
I^A_k &=& 
\Big(\frac{1}{t}-t\Big)^k \prod_{I=1}^k \frac{dz_I}{z_I}
 \prod_{I<J}^k 
 \frac{\left(1 -  \left(\frac{z_I}{z_J}\right)^{\pm 1}\right) \left(1- \left(\frac{z_I}{z_J}\right)^{\pm 1}t^2\right)}{t^2}
\\[10pt]
I^q_k &=& \prod_{I=1}^k \prod_{i=1}^N
\frac{t}{\left(1-\left(\frac{z_I}{s_i}\right)^{\pm 1}t\right)}
\\[10pt]
I^X_k &=& \prod_{I=1}^k \frac{t}{(1- u^{\pm 1}t)}
\prod_{I<J}^k \frac{t^2}{\left(1-\left(\frac{z_I}{z_J}\right)^{\pm 1} u^{\pm 1} t\right)}
\\[10pt]
I^Y_k &=& \prod_{I=1}^k \frac{t^2}{(1-(az_I^2)^{\pm 1} v^{\pm 1} t)}
\prod_{I<J}^k \frac{t^2}{(1-(az_I z_J)^{\pm 1} v^{\pm 1} t)}\\[10pt]
I^{\psi_R}_k &=& \prod_{I=1}^k \prod_{i=1}^N
\frac{(1 - (az_I s_i)^{\pm 1} v)}{v}\\[10pt]
I^{\chi_R}_k &=& \prod_{I<J}^k \frac{(1-(az_I z_J)^{\pm 1} v^{\pm 1} u)}{u^2}
\\[10pt]
I^{\xi_R}_k &=& \prod_{I=1}^k \prod_{a=1}^{N_f} (\sqrt{z_I f_a} - \frac{1}{\sqrt{z_I f_a}})
\ee
We have included here the fugacity $a$ of the additional $U(1)_{\ell'}$ symmetry.

\subsection{$Sp(N_1)\times Sp(N_2) + (\funda,\funda)$}

The content of the 1d theory in this case is shown in Fig.~\ref{SpSpwithFlavors}.
The contributions of $A$, $X$, $q$, $\psi_R$ and $\xi_R$ are essentially the same as in the $Sp(N) + \asymm$ theory, 
with $N$ replaced by either $N_1$ or $N_2$, and $n$ replaced by either $n_1$ or $n_2$, as can be read off directly from the quiver diagram.
The remaining contributions of $Y$ and $\chi_R$ are given for 
$k_1=2n_1$ and $k_2=2n_2$ by:
\be
I_{k_1,k_2}^{Y,\epsilon_1,\epsilon_2} = \left\{
\begin{array}{ll}
{\displaystyle\prod_{I,J=1}^{n_1,n_2}} \frac{t^4}{(1- z_{1I}^{\pm 1} z_{2J}^{\pm 1} v^{\pm 1} t)} & (\epsilon_1,\epsilon_2) = (+,+) \\
{\displaystyle\prod_{I=1}^{n_1}}  \frac{t^4}{(1 - z_{1I}^{\pm 2} v^{\pm 2}t^2)}
{\displaystyle\prod_{I,J=1}^{n_1,n_2-1}} \frac{t^4}{(1- z_{1I}^{\pm 1} z_{2J}^{\pm 1} v^{\pm 1} t)}
& (\epsilon_1,\epsilon_2) = (+,-) \\
{\displaystyle\prod_{J=1}^{n_2}}  \frac{t^4}{(1 - z_{2J}^{\pm 2} v^{\pm 2}t^2)}
{\displaystyle\prod_{I,J=1}^{n_1-1,n_2}} \frac{t^4}{(1- z_{1I}^{\pm 1} z_{2J}^{\pm 1} v^{\pm 1} t)}
& (\epsilon_1,\epsilon_2) = (-,+) \\
\frac{t^2}{(1- v^{\pm 2}t^2)}
{\displaystyle\prod_{I=1}^{n_1-1}} \frac{t^4}{(1 - z_{1I}^{\pm 2} v^{\pm 2}t^2)}
{\displaystyle\prod_{J=1}^{n_2-1}} \frac{t^4}{(1 - z_{2J}^{\pm 2} v^{\pm 2}t^2)}
{\displaystyle\prod_{I,J=1}^{n_1-1,n_2-1}} \frac{t^4}{(1- z_{1I}^{\pm 1} z_{2J}^{\pm 1} v^{\pm 1} t)}
&  (\epsilon_1,\epsilon_2) = (-,-)
 \end{array}
 \right.\\[10pt]
I_{k_1,k_2}^{\chi_R,\epsilon_1,\epsilon_2} = \left\{
\begin{array}{ll}
{\displaystyle\prod_{I,J=1}^{n_1,n_2}} \frac{(1- z_{1I}^{\pm 1} z_{2J}^{\pm 1} v^{\pm 1} u)}{u^4} & (\epsilon_1,\epsilon_2) = (+,+) \\
{\displaystyle\prod_{I=1}^{n_1}}  \frac{(1 - z_{1I}^{\pm 2} v^{\pm 2}u^2)}{u^4}
{\displaystyle\prod_{I,J=1}^{n_1,n_2-1}} \frac{(1- z_{1I}^{\pm 1} z_{2J}^{\pm 1} v^{\pm 1} u)}{u^4}
& (\epsilon_1,\epsilon_2) = (+,-) \\
{\displaystyle\prod_{J=1}^{n_2}}  \frac{(1 - z_{2J}^{\pm 2} v^{\pm 2}u^2)}{u^4}
{\displaystyle\prod_{I,J=1}^{n_1-1,n_2}} \frac{(1- z_{1I}^{\pm 1} z_{2J}^{\pm 1} v^{\pm 1} u)}{u^4}
& (\epsilon_1,\epsilon_2) = (-,+) \\
\frac{(1- v^{\pm 2}u^2)}{u^2}
{\displaystyle\prod_{I=1}^{n_1-1}} \frac{(1 - z_{1I}^{\pm 2} v^{\pm 2}u^2)}{u^4}
{\displaystyle\prod_{J=1}^{n_2-1}} \frac{(1 - z_{2J}^{\pm 2} v^{\pm 2}u^2)}{u^4}
{\displaystyle\prod_{I,J=1}^{n_1-1,n_2-1}} \frac{(1- z_{1I}^{\pm 1} z_{2J}^{\pm 1} v^{\pm 1} u)}{u^4}
&  (\epsilon_1,\epsilon_2) = (-,-)
 \end{array}
 \right.
\ee
and for $k_1=2n_1 + 1$ and $k_2=2n_2+1$ by:
\be\nonumber
I_{k_1,k_2}^{Y,\epsilon_1,\epsilon_2}&=& 
\frac{t}{(1-\epsilon_1\epsilon_2 v^{\pm 1}t)}
\prod_{I=1}^{n_1} \frac{t^2}{(1 - \epsilon_2 z_{1I}^{\pm 1} v^{\pm 1}t)}
\prod_{J=1}^{n_2} \frac{t^2}{(1 - \epsilon_1 z_{2J}^{\pm 1} v^{\pm 1}t)}
\prod_{I,J=1}^{n_1,n_2} \frac{t^4}{(1- z_{1I}^{\pm 1} z_{2J}^{\pm 1} v^{\pm 1} t)}\\
\ee
\be\nonumber
I_{k_1,k_2}^{\chi_R,\epsilon_1,\epsilon_2} &=& 
\frac{(1-\epsilon_1\epsilon_2 v^{\pm 1}u)}{u}
\prod_{I=1}^{n_1} \frac{(1 - \epsilon_2 z_{1I}^{\pm 1} v^{\pm 1}u)}{u^2}
\prod_{J=1}^{n_2} \frac{(1 - \epsilon_1 z_{2J}^{\pm 1} v^{\pm 1}u)}{u^2}
\prod_{I,J=1}^{n_1,n_2} \frac{(1- z_{1I}^{\pm 1} z_{2J}^{\pm 1} v^{\pm 1} u)}{u^4} \,,
\ee
and for $k_1 = 2n_1+1$ and $k_2=2n_2$ by:
\be
I_{k_1,k_2}^{Y,\epsilon_1,\epsilon_2} &=& \left\{
\begin{array}{ll}
{\displaystyle\prod_{J=1}^{n_2}} \frac{t^2}{(1 - \epsilon_1 z_{2J}^{\pm 1} v^{\pm 1}t)}
{\displaystyle\prod_{I,J=1}^{n_1,n_2}} \frac{t^4}{(1- z_{1I}^{\pm 1} z_{2J}^{\pm 1} v^{\pm 1} t)} & \epsilon_2 = + \\
\frac{t^2}{(1- v^{\pm 2}t^2)}
{\displaystyle\prod_{J=1}^{n_2-1}} \frac{t^2}{(1 - \epsilon_1 z_{2J}^{\pm 1} v^{\pm 1}t)}
{\displaystyle\prod_{I=1}^{n_1}} \frac{t^4}{(1 - z_{1I}^{\pm 2} v^{\pm 2}t^2)}
{\displaystyle\prod_{I,J=1}^{n_1,n_2-1}} \frac{t^4}{(1- z_{1I}^{\pm 1} z_{2J}^{\pm 1} v^{\pm 1} t)}
& \epsilon_2 = -
 \end{array}
 \right.
\\[10pt]
I_{k_1,k_2}^{\chi_R,\epsilon_1,\epsilon_2} &=& \left\{
\begin{array}{ll}
{\displaystyle\prod_{J=1}^{n_2}} \frac{(1 - \epsilon_1 z_{2J}^{\pm 1} v^{\pm 1}u)}{u^2}
{\displaystyle\prod_{I,J=1}^{n_1,n_2}} \frac{(1- z_{1I}^{\pm 1} z_{2J}^{\pm 1} v^{\pm 1} u)}{u^4} & \epsilon_2 = + \\
\frac{(1- v^{\pm 2}u^2)}{u^2}
{\displaystyle\prod_{J=1}^{n_2-1}} \frac{(1 - \epsilon_1 z_{2J}^{\pm 1} v^{\pm 1}u)}{u^2}
{\displaystyle\prod_{I=1}^{n_1}} \frac{(1 - z_{1I}^{\pm 2} v^{\pm 2}u^2)}{u^4}
{\displaystyle\prod_{I,J=1}^{n_1,n_2-1}} \frac{(1- z_{1I}^{\pm 1} z_{2J}^{\pm 1} v^{\pm 1} u)}{u^4}
& \epsilon_2 = -
 \end{array}
 \right.
\ee
The case $k_1=2n_1$ and $k_2=2n_2+1$ is related to the last one by replacing $n_1 \leftrightarrow n_2$, $z_{1I} \leftrightarrow z_{2J}$
and $\epsilon_1 \leftrightarrow \epsilon_2$.

\section{$SU(2)\times Sp(2)$ in the $SU(N)$ formalism}

The $SU(N)$ instanton partition function is generically extracted from the $U(N)$ one by imposing the constraint $\prod_{i=1}^N s_i = 1$.
However there are cases where a correction factor must be included to remove the contribution of another type of extraneous states.
In 5-brane constructions these correspond to compact string states that are external to the 5-brane web \cite{Bergman:2014kza}.
The general structure of the factor is 
\be
Z_{extra} = PE\Bigg[\frac{t^2 \mathcal{F}}{(1-tu)(1-\frac{t}{u})}\Bigg] \,,
\ee
where $\mathcal{F}$ depends the instanton and global symmetry fugacities, and is determined on a case by case basis,
typically by requiring the invariance of the full partition function under $t\rightarrow 1/t$, which is a part of the conformal symmetry.
For $SU(2)\times Sp(2)$ we find that ${\cal F} = (v^2 + v^{-2})q_1$, and therefore
 \be
\label{11SU2Sp2}
Z^{SU(2)_{\theta_1}\times Sp(2)_{\theta_2}}=PE\Bigg[\frac{t^2 (v^2+\frac{1}{v^2})q_1}{(1-tu)(1-\frac{t}{u})}\Bigg]Z^{U(2)_{\theta_1/\pi}\times Sp(2)_{\theta_2}}/
\{s_1s_2=1\} \,.
\ee
This can now be computed by incorporating the results of \cite{Bergman:2014kza} for $U(N)\times Sp(N)$.
We will not do this explicitly here, but rest assured that the result at order $q_1 q_2$ agrees with (\ref{Sp1Sp2}).

\end{appendices}

\end{document}